\documentclass[showpacs,preprintnumbers,
superscriptaddress,amsmath,floatfix,prd]{revtex4}

\usepackage{graphicx}

\usepackage[dvips]{color}

\usepackage[normalem]{ulem}  % \sout{old text} for strikeout

\renewcommand\a{\alpha}
\renewcommand\b{\beta}
\newcommand\g{\gamma}
\renewcommand\d{\delta}
\newcommand\e{\epsilon}
\newcommand\ve{\varepsilon}

\renewcommand\l{\lambda}
\newcommand\m{\mu}
\newcommand\n{\nu}

\newcommand\G{\Gamma}

\newcommand{\non}{\nonumber\\}

\newcommand{\be}{\begin{equation}}
\newcommand{\ee}{\end{equation}}
\newcommand{\bea}{\begin{eqnarray}}
\newcommand{\eea}{\end{eqnarray}}
\newcommand{\ba}[1]{\begin{array}{#1}}
\newcommand{\ea}{\end{array}}

\newcommand{\up}{\hat{\mathbf{p}}}
\newcommand{\hk}{\hat{k}}

\newcommand{\hp}{\hat{p}}
\newcommand{\vg}{\bm{\gamma}}

\newcommand{\uk}{\hat{\mathbf{k}}}

\newcommand{\bm}[1]{\mbox{\boldmath${#1}$}}

\begin{document}

\title{Bulk viscosity in 2SC quark matter}

\author{Mark G.\ Alford}
\email{alford@wuphys.wustl.edu}
\affiliation{Department of Physics, Washington University St Louis, MO, 63130, USA}

\author{Andreas Schmitt}
\email{aschmitt@wuphys.wustl.edu}
\affiliation{Department of Physics, Washington University St Louis, MO, 63130, USA}

\date{November 2, 2006}

\begin{abstract}

The bulk viscosity of three-flavor color-superconducting quark matter
originating from the nonleptonic process $u+s\leftrightarrow u+d$ is
computed. It is assumed that up and down quarks form Cooper pairs
while the strange quark remains unpaired (2SC phase). A general
derivation of the rate of strangeness production is presented,
involving contributions from a multitude of different subprocesses,
including subprocesses that involve different numbers of gapped quarks
as well as creation and annihilation of particles in the condensate.
The rate is then used to compute the bulk viscosity as a function of
the temperature, for an external oscillation frequency typical of a
compact star $r$-mode.  We find that, for temperatures far below the critical
temperature $T_c$ for 2SC pairing, the bulk viscosity of
color-superconducting quark matter is suppressed relative to that of
unpaired quark matter, but for $T\gtrsim 10^{-3}\,T_c$ the
color-superconducting quark matter has a higher bulk viscosity.
This is potentially relevant for the suppression of $r$-mode instabilities 
early in the life of a compact star.

\end{abstract}


\maketitle

\section{Introduction}
\label{intro}

Compact stars consist of extremely dense matter, part or all of which
may be deconfined quark matter \cite{perry}. Quark matter that is sufficiently
dense and cold is expected to be in a color-superconducting state \cite{bailin},
characterized by Cooper pairing between the quarks, mediated by their
attractive strong interaction. This is analogous to the formation of
electron Cooper pairs in a superconducting metal \cite{bcs}. 
However, unlike electrons, quarks occur in many varieties: there are three
colors and in a compact star core we expect to find three flavors, so a
multitude of pairing (and thus symmetry breaking) patterns and hence
different phases is theoretically possible \cite{reviews}. It is an
open problem of current research to determine the ground state out of
these many possibilities for quark matter inside a compact star.

One approach to this problem is purely theoretical, using perturbative QCD as well as more phenomenological 
models in order to compare different possibilities and find the one with the lowest free energy. 
This approach yields a definite answer for very large densities (more precisely for quark chemical 
potentials $\mu$ much larger than the masses of the up, down, and strange quarks). 
In this situation, quark matter 
is in the color-flavor locked (CFL) state \cite{Alford:1998mk}. However, in nature, i.e., in the interior
of a compact star, the quark chemical potential is expected to be of the order of 500 MeV or less, 
which requires the mass of the strange quark $m_s$ to be taken into account. Moreover, further 
constraints on the matter are imposed by electric and color charge neutrality as well as by equilibrium
with respect to the weak interaction. All these conditions impose a stress on the pairing, no matter 
which pairing pattern is chosen \cite{Rajagopal:2005dg}. Therefore, besides the CFL phase, many 
unconventional color-superconducting phases have been proposed, e.g., phases that break rotational
\cite{spin1} or translational \cite{LOFF} symmetries.   
  
The second approach to determining the ground state of dense quark
matter in a compact star is the study of possible astrophysical
effects of color superconductors. In particular, the ultimate goal is
to compare observational data with theoretical predictions for color
superconductors in order to exclude or confirm proposed
phases. Moreover, one would like to use observations to distinguish
between quark matter and more conventional matter inside the star,
such as nuclear or hyperonic matter.  Promising physical properties that
are distinct for different color-superconducting phases are related to
transport properties such as thermal conductivity
\cite{Shovkovy:2002kv}, neutrino emissivity
\cite{Carter:2000xf,Jaikumar:2005hy,Schmitt:2005wg,Wang:2006tg},
magnetic fields \cite{Alford:1999pb,Ferrer:2005vd}, and viscosity.

In this paper, we calculate the bulk viscosity of
color-superconducting quark matter in the 2SC phase. (For a recent
study of the bulk viscosity in spin-one color superconductors, see
Ref.~\cite{Sa'd:2006qv}; for hybrid/CFL matter see Ref.\ \cite{Drago:2003wg}.
).  In 2SC quark matter, up and down quarks
form Cooper pairs with each other while the strange quarks remain
unpaired.
Although constraints of charge neutrality tend to disfavor
the 2SC phase \cite{Alford:2002kj,Steiner:2002gx}, 
it is found in phase diagrams calculated from NJL models,
for example in Ref.~\cite{Ruster:2005jc} 
%for stronger diquark coupling ($G_D=G_S$)
%at $T=0$, over the chemical potential range 
%$398~\MeV \lesssim \mu \lesssim 412~\MeV$
(see their Figs.~2 and 6) and in Ref.~\cite{Abuki:2005ms}
(see their Fig.~7). In our calculations we shall neglect the effect of 
electric charge neutrality, so we treat the up and down quarks as having the
same chemical potential and hence the same Fermi momentum.
This has a quantitative effect, which we expect to
be of order $m_s^2/\mu^2$,
on the available phase space for the quark momenta. 
We therefore expect our results to deviate only 
slightly from the results for electrically neutral 2SC matter. 

Bulk viscosity, as well as shear viscosity
\cite{Manuel:2004iv}, is relevant to vibrational and rotational modes
of a compact star, whose frequencies are typically in the kHz range
\cite{Kokkotas:2001ze}.
In particular, for a rapidly rotating star there is the possibility
of rotational ``$r$-modes'' 
that are unstable with respect to gravitational
radiation \cite{Andersson:1997xt,Andersson:2001bz,Lindblom:2000jw,Weber:2004kj}.
This would lead to a drastic decrease of the rotation frequency of the
star, which is incompatible with observations of rapidly rotating
pulsars with periods of the order of ms. One mechanism to damp the
instabilities is a nonzero viscosity of the fluid \cite{cutler}.

In general, bulk viscosity characterizes the response of the system to an
externally imposed oscillating volume change.  If the volume
compression (and expansion) induces a change in the chemical
composition of the matter, i.e., if it drives the system out of
chemical equilibrium, then microscopic processes will try to
re-equilibrate the system. If these processes occur on a timescale
comparable to that of the driving oscillation then power dissipation
will occur.
% Bulk viscosity is determined given by the power dissipation that
% occurs due to the out-of-phase oscillations of the volume and the
% chemical reequilibration processes.
Because the strange quark is massive, a
volume oscillation causes a difference between the chemical potential
of the $s$ quark and that of the $u$ and $d$ quarks.  The relevant
microscopic re-equilibration process is the weak process
$u+d\leftrightarrow u+s$ \cite{Wang:1985tg} which produces (annihilates)
strangeness. (Leptonic processes are suppressed by their smaller phase
space. However, we shall discuss the possibility
that they might still yield a significant contribution to
the bulk viscosity.) Strong processes can be neglected
since they occur on much shorter time scales than the scale given by the oscillation 
of the star. This is in contrast to the context of heavy ion collisions where gluonic
processes are important for the bulk viscosity \cite{Arnold:2006fz}. Moreover, we do 
not consider a potential contribution from the Goldstone boson associated with the 
spontaneous breaking of $U(1)_A$ because this symmetry is explicitly broken at densities
relevant for compact stars. Other (approximate) global symmetries are not broken in the 2SC phase, 
hence there are no other (pseudo) Goldstone bosons. 

In order to compute the rate of the process $u+d\leftrightarrow u+s$ 
one has to consider the corresponding collision integral. 
In the case of a superconductor/superfluid this integral is
considerably more complicated than in the unpaired phase. Some of the
complication originates from the broken number conservation, allowing
not only processes that consist of two ingoing and outgoing particles
but also processes with three (four) particles coalescing into one
(zero) particles and one (zero) particles decaying into three (four)
particles. In these processes, particles are created from or deposited
into the condensate. This effect is also present in more conventional
superfluids such as $^3$He \cite{bhatta,vollhardt}. Starting from the kinetic
equation, we present a detailed derivation of the collision integral
which contains all these possible processes.

An additional complication arises from the nature of the 2SC
phase. The process $u_1+d\leftrightarrow u_2+s$ actually contains
several subprocesses according to whether the incoming and outgoing
states are gapped (up or down quarks that are red or green) or ungapped
(all the other species).  All these subprocesses are included
naturally in our formal derivation. We encounter processes where zero,
one, two, or three of the quarks $u_1$, $d$, $u_2$, $s$ are
gapped. Therefore, the 2SC phase is an interesting system to study the
more general question of the quantitative effects of the energy gap
for the rate of the microscopic process. We compute the rates for
arbitrary temperatures $T$ (up to the superconducting transition
temperature), i.e., we go beyond the low-temperature regime, where the
effect of the energy gap $\Delta$ is an exponential suppression of the
rate. 

A final subtlety is that the bulk viscosity is not a monotonic
function of the rates of the equilibrating processes. The bulk
viscosity is highest when those rates are closest to the time scale of
the oscillation that is being damped. As we will see, this means that,
depending what frequencies are physically relevant,
a phase in which equilibration proceeds more slowly may have
a higher bulk viscosity than one in which it proceeds quickly.
Thus the 2SC phase, whose equilibration rate is
always slower than that of unpaired quark matter, will turn out to
have a larger bulk viscosity at kHz frequencies
over a wide range of temperatures.

The paper is organized as follows. In Sec.~\ref{definebulk} we define the bulk viscosity and 
point out an analogy to the power dissipation in an electric circuit with a capacitance and a 
resistance. The derivation of the rate of the process $u+d\leftrightarrow u+s$ in 2SC quark matter 
is presented in Sec.~\ref{deriverate}. At the end of this section, in Sec.~\ref{subprocesses}
we discuss the physical content of the derived expression. In Sec.~\ref{evaluaterate}, the rate is evaluated 
in full generality and several limit cases (unpaired phase, zero-temperature limit) as well 
as several subprocesses and their contribution to the full result is discussed. We use these results
to compute the bulk viscosity in Sec.~\ref{calcbulk} and conclude with Sec.~\ref{conclusions}.

\section{Definition of bulk viscosity}
\label{definebulk}

In this section we derive the expression for the bulk viscosity with respect to the 
process $u+s\leftrightarrow u+d$. This derivation is based on respective derivations of unpaired 
quark matter and variations of it for different microscopic 
processes \cite{madsen,anand,Haensel:2000vz,Lindblom:2001hd}. We also point out an instructive analogy 
to an electric circuit with resistance and capacitance.

Before we do so, we notice that the following derivation includes several assumptions
about other processes that potentially contribute to the bulk viscosity. First, there may be
a contribution from the leptonic processes $u+e\leftrightarrow d + \nu$ and $u+e\leftrightarrow s + \nu$. 
In Appendix \ref{appleptonic} we present a derivation of the bulk viscosity including these
processes in addition to the nonleptonic process $u+s\leftrightarrow u+d$. The result (\ref{multibulk})
is complicated compared to the result (\ref{bulkfinal}), derived in this section. 
Because of phase space restrictions,
the leptonic processes can be expected to be slower than the non-leptonic one. However, as we 
discuss in Appendix \ref{appleptonic}, their contribution can still be significant, depending 
on the external oscillation frequency. In this paper, we do not compute the rates of the leptonic
processes microscopically. Therefore, we leave the study of their quantitative contribution to the bulk 
viscosity for future work.

Second, the derivation of the bulk viscosity below treats the effect of the process 
$u+s\leftrightarrow u+d$ solely through the flavor chemical 
potentials $\mu_u$, $\mu_d$, and $\mu_s$, disregarding color degrees of freedom. This treatment 
is correct under the assumption of infinitely fast, color-changing strong processes as we 
show in Appendix \ref{appstrong}. In this appendix we set up a more general derivation of the 
bulk viscosity showing that in the absence of the strong interaction (or for not infinitely fast 
strong processes) the expression for the bulk viscosity is more complicated.
For this aspect, see also discussion of the results of Fig.~\ref{figbulk0123}.

We start from the definition of the bulk viscosity \cite{madsen,anand,Haensel:2000vz,Lindblom:2001hd} 
\be \label{defbulk}
\zeta \equiv 2\left\langle\frac{dW}{dt}\right\rangle\left(\frac{V_0}{\delta V_0}\right)^2\frac{1}{\omega^2} \, ,
\ee
where we assume the volume of the system to oscillate with angular frequency $\omega$ and amplitude 
$\delta V_0$ about its equilibrium value $V_0$, 
\be \label{vosc}
V(t)=V_0+\d V(t)\, , \qquad \d V(t) = \d V_0\cos\omega t \, .
\ee
The average dissipated power per 
volume in one oscillation period $\tau\equiv 2\pi/\omega$ is 
\be \label{power}
\left\langle\frac{dW}{dt}\right\rangle = - \frac{1}{\tau\,V_0}\int_0^\tau p(t)\frac{dV}{dt}dt \, ,
\ee
where $p(t)$ is the pressure.
We shall express the dissipated power in terms of $\d V(t)$ and the resulting oscillation in 
$\d\m(t)$. Here, $\d\m$ is the difference between the $s$ quark and the $d$ quark 
chemical potential $\d\m=\mu_s-\mu_d$ which vanishes in chemical equilibrium. 
To this end, we write the pressure for small oscillations as 
\be \label{pressure}
p(t) = p_0 + \left(\frac{\partial p}{\partial V}\right)_0 \d V(t)+ 
\left(\frac{\partial p}{\partial n_d}\right)_0 \d n_d(t)+ \left(\frac{\partial p}{\partial n_s}\right)_0 
\d n_s(t)
\, .
\ee
Here and in the following, the subscript 0 denotes equilibrium, i.e., $\delta\mu=\delta V =0$, and $n_s$ and 
$n_d$ are the $s$- and $d$-quark number densities, respectively. We employ the thermodynamic relation
\be \label{p1}
\left(\frac{\partial p}{\partial n_i}\right)_0 = - V_0 \left(\frac{\partial \mu_i}{\partial V}\right)_0 
\, , \qquad i=d,s \, , 
\ee
and use
\begin{subequations} \label{derivs}
\bea
- V_0 \left(\frac{\partial \mu_d}{\partial V}\right)_0&=& 
\left(n_d\frac{\partial \mu_d}{\partial n_d}\right)_0 + 
\left(n_u\frac{\partial \mu_d}{\partial n_u}\right)_0   \, , \\
- V_0 \left(\frac{\partial \mu_s}{\partial V}\right)_0&=& 
\left(n_s\frac{\partial \mu_s}{\partial n_s}\right)_0   \, . 
\eea
\end{subequations}
Here we have taken into account that, in the 2SC phase, $u$ and $d$ quarks form Cooper pairs with 
each other. Therefore, $\mu_u$ and $\mu_d$ depend on {\em both} $n_u$ and $n_d$. In contrast,
$\mu_s$ solely depends on $n_s$ because $s$ quarks are unpaired. See Table \ref{tableBC} for 
the explicit form of the densities.
Next, we express the change in densities in terms of the rate of the microscopic process, to be computed in 
Secs.~\ref{deriverate} and \ref{evaluaterate},
\be \label{p2}
\d n_d(t) = - \d n_s(t) = \int_0^t \Gamma_d[\d\m(t')]\,dt' \, . 
\ee
Here, $\Gamma_d[\d\m(t)]$ is the number of $d$ quarks per volume and time produced in the reaction
$u+s\leftrightarrow u+d$. Since both directions of the reaction are taken into account, the rate is zero 
for $\d\m=0$. In the present situation, however, a nonzero $\d\m$ is induced through the volume oscillation,
rendering $\Gamma_d[\d\m(t)]$ nonzero. We shall use the linear approximation
\be \label{deflambda}
\Gamma_d[\d\m(t)] = \lambda\, \d\m(t) \,  .
\ee 
Inserting Eqs.\ (\ref{derivs}) into (\ref{p1}), the result and (\ref{p2}) into Eq.\ (\ref{pressure}), 
the result into Eq.\ (\ref{power}), and making use of Eq.\ (\ref{deflambda}) we find
\be \label{resultpower}
\left\langle\frac{dW}{dt}\right\rangle = -\frac{B}{V_0}\left\langle\left(\int_0^t 
\Gamma_d[\d\m(t')]\,dt'\right) \frac{dV}{dt}\right\rangle 
= B\lambda\,\langle\d\m(t)\,\d v(t)\rangle
\, ,
\ee
where the right-hand side has been obtained via partial integration, we have abbreviated
$\d v(t)\equiv \d V(t)/V_0$, and 
\be \label{defB}
B\equiv \left(n_u\frac{\partial \mu_d}{\partial n_u}\right)_0+
\left(n_d\frac{\partial \mu_d}{\partial n_d}\right)_0-\left(n_s\frac{\partial \mu_s}{\partial n_s}
\right)_0 \,\, . 
\ee
Finally, we have to connect the change in chemical potentials to the volume oscillation and 
the microscopic rate, 
\be \label{deltamu}
\frac{\partial \d\mu}{\partial t} = \left(\frac{\partial \d\m}{\partial V}\right)_0 \frac{dV}{dt}
+\left(\frac{\partial \d\m}{\partial n_d}\right)_0 \frac{dn_d}{dt} +  
\left(\frac{\partial \d\m}{\partial n_s}\right)_0 \frac{dn_s}{dt} \, , 
\ee
where the change in densities is due to the microscopic process, $dn_d/dt=-dn_s/dt=\Gamma_d$. Consequently,
\be \label{diffeq}
\frac{\partial \d\mu}{\partial t}
= B\,\frac{\partial \delta v}{\partial t} - \gamma \,\d\mu(t)  \, , 
\ee
where we abbreviated the characteristic (inverse) time scale set by the microscopic process by
\be \label{defgamma}
\g\equiv C\lambda 
\ee
where
\be \label{defC}
C\equiv \left(\frac{\partial \mu_d}{\partial n_d}\right)_0
+\left(\frac{\partial \mu_s}{\partial n_s}\right)_0 \, .
\ee
The equation for the power dissipation (\ref{resultpower}) in terms of the periodic external quantity 
$\d v(t)$ and the response of the system $\d\m(t)$ together with the differential equation (\ref{diffeq})
for $\d\m(t)$ can be mapped exactly onto the analogous situation in an electric circuit. In this
case, the external quantity is an alternating voltage $U(t)$ while the response of the system 
is the current $I(t)$. For a circuit with a resistance $R$, an inductance $L$ and
a capacitance $C'$, the corresponding equations are 
\begin{subequations}
\bea
\left\langle\frac{dW}{dt}\right\rangle &=& \langle I(t)\, U(t)\rangle \, , \\
\dot{U}(t) &=& \dot{I}(t) R + L \ddot{I}(t) + \frac{I(t)}{C'} \, . \label{voltage}
\eea
\end{subequations}
Obviously, Eq.\ (\ref{voltage}) has the same form as Eq.\ (\ref{diffeq}) and we can identify
``voltage'', ``current'', ``resistance'', ``inductance'', and ``capacitance'' in the 
color superconductor,
\be \label{replace}
U(t) \to \d v(t) \, , \qquad I(t) \to \d\m(t) \, , \qquad R\to \frac{1}{B}\, , 
\qquad L \to 0 \, , \qquad C' \to \frac{B}{\gamma}   \, .
\ee
This analogy with an electric circuit is not only of a mathematical nature, we can also use it to 
understand the underlying physics. This is useful because the behavior of the bulk viscosity, 
unlike the shear viscosity, is somewhat counter-intuitive if one applies the daily life picture of a 
``viscous'' fluid. 
Therefore, let us interpret the analogies in Eq.\ (\ref{replace}). First, the ``resistance'' of the system
is proportional to $1/B$ with $B$ given in Eq.\ (\ref{defB}). In particular, the resistance is infinite 
for vanishing $B$. This means that there is no response of the system to the external force, just as there 
is no current in a circuit with infinite resistance. The reason is that a vanishing $B$ means that  
there is no difference of $d$ and $s$ flavor in the response to the change in density induced by the 
volume oscillations. Hence the system is still in chemical equilibrium and no energy is 
dissipated.
A finite resistance occurs if there is a difference in the dispersion relations of $d$ ans $s$ quarks, 
such as a different mass or different energy gaps. Only then does the system respond with an oscillating
$\d\m(t)$ and both the power dissipation and the bulk viscosity are non-vanishing. 

The ``capacitance'' is
proportional to $1/\gamma$, where $\gamma$ is the inverse time scale that characterizes the weak process. 
A slow process has a small $\gamma$, hence a large capacitance, meaning that the system can store 
a large amount of energy. This energy is only slowly released by producing $d$ quarks 
(or $s$ quarks, depending on the sign of $\d\m$) and thus re-approaching chemical equilibrium. This is 
just like a large capacitance which can store a large amount of electrical charge. 
On the other hand, a fast process has a large $\gamma$ and the capacitor can be discharged quickly
(i.e.~chemical equilibrium can be reached quickly). 

Let us now continue with our derivation. 
We write the response of the system as
\be 
\d\m(t) = {\rm Re}(\delta\mu_0\, e^{i\omega t}) 
\ee
with a complex amplitude $\delta\m_0$. This expression is inserted into the differential equation
(\ref{diffeq}). We obtain a phase lag $\phi$ of $\d\m(t)$ with respect to 
$\d v(t)$, 
\be 
\phi=\arctan\frac{\g}{\omega} \, .
\ee
The amplitude of the oscillation of the chemical potential is
\be \label{amplitude}
|\delta\m_0| =  \frac{\delta V_0}{V_0}\frac{B\omega }{\sqrt{\g^2+\omega^2}} \, ,
\ee
and the power dissipation is
\be
\left\langle\frac{dW}{dt}\right\rangle = \frac{1}{2}\, \frac{\delta V_0}{V_0}\,B\,\lambda\,
{\rm Re}\,\delta\mu_0 = 
\frac{1}{2}\left(\frac{\delta V_0}{V_0}\right)^2\frac{B^2\lambda\,\omega^2}
{\g^2+\omega^2}
\, .
\ee
With the definition (\ref{defbulk}) we arrive at the following expression for the bulk viscosity,
\be \label{bulkfinal}
\zeta = \frac{B^2}{C}\frac{\gamma}{\gamma^2+\omega^2} \, , 
\ee
This expression, together with the definitions (\ref{defB}), (\ref{defgamma}), and (\ref{defC}), 
is the main result of this section and shall be used in Sec.~\ref{calcbulk}.

In Table \ref{tableBC}, we show the explicit values for the coefficients $B$ and $C$ in unpaired
and 2SC quark matter at zero temperature, derived from the quark densities as functions 
of the chemical potentials. The densities can be computed from the respective free energy in 
Ref.~\cite{Alford:2002kj}. In order to obtain the partial derivatives appearing in 
$B$ and $C$ in the 2SC phase, one has to 
invert the Jacobian of the two-dimensional function $(n_u,n_d)=(n_u(\mu_u,\mu_d),n_d(\mu_u,\mu_d))$.
We have neglected terms in higher orders of $m_s$ and the pairing gap $\Delta$.
Moreover, we expect the difference in chemical potentials $\d\n\equiv(\mu_u-\mu_d)/2$ to be of the order
of $m_s^2/\mu$ \cite{Alford:2002kj}. Therefore, we have also dropped higher order terms in $\d\n$. 
In the following calculation of the microscopic rate, we shall, for simplicity, consider the 
limit $\mu_u=\mu_d$. It is important to note that even in the limit $\Delta=0$ the value of the
coefficient $C$ is different in the two considered phases. The reason is the ``locking''
of chemical potentials of the pairing fermions. In other words, gapped quasiparticles in the
2SC phase look as if their chemical potential (in the absence of pairing) was $\n\equiv (\mu_u+\mu_d)/2$.
This effect, the difference in $C$ for unpaired and 2SC matter, is weakened by a nonzero temperature.
We have checked numerically, however, that the difference remains sizable for all temperatures
up to the transition temperature. Thus, for simplicity, we shall use the zero-temperature values
from Table \ref{tableBC} as an approximation. 

We have checked that electric neutrality has no sizable effect on the coefficients $B$ and $C$. 
Requiring local electric neutrality at all times
yields a constraint for the densities 
and chemical potentials via the condition $2n_u-n_d-n_s=0$. Implementing this condition
modifies the derivation for the bulk viscosity. However, to leading order the value of $B^2/C$ turns out to be the same as in the non-neutral case.  

\begin{table}  
\begin{tabular}[t]{|c||c|c|}
\hline
& unpaired & 2SC \\ \hline\hline
$\;\;n_u\;\;$ & $\mu_u^3/\pi^2$ & 
$(\mu_u^3/3 + 2\nu^3/3 + \Delta^2\nu)/\pi^2$\\ \hline
$n_d$ & $\mu_d^3/\pi^2$& 
$(\mu_d^3/3 + 2\nu^3/3 + \Delta^2\nu)/\pi^2$\\ \hline
$n_s$ & $\mu_s^3/\pi^2\;[1-3m_s^2/(2\mu^2_s)]$& 
$\mu_s^3/\pi^2\;[1-3m_s/(2\mu^2_s)]$\\ \hline
$B$ & $m_s^2/(3\mu_d)$& $m_s^2/(3\mu_d)+2\Delta^2/(9\nu)$ \\ \hline
$C$ & $\;\;2\pi^2/(3\mu_d^2)\;[1+m_s^2/(4\mu_d^2)]\;\;$ & 
$\;\;\pi^2/\mu_d^2\;[
1+m_s^2/(6\mu_d^2)-\Delta^2/(18\mu_d^2)-2\d\n/(3\mu_d)]\;\;$ \\ \hline
$B^2/C$ & $m_s^4/(6\pi^2)$ & $m_s^4/(9\pi^2) + 4\Delta^2m_s^2/(27\pi^2) +4\Delta^4/(81\pi^2)$ \\
\hline
\end{tabular}
\caption{
Quark densities $n_u$, $n_d$, $n_s$ and coefficients $B$ and $C$ for
unpaired and 2SC quark matter for massless $u$ and $d$ quarks and a
strange mass $m_s\ll \mu_u,\mu_d,\mu_s$. We have abbreviated
$\nu\equiv (\mu_u+\mu_d)/2$ and $\d\nu\equiv (\mu_u-\mu_d)/2$. The
energy gap due to pairing of $u$ and $d$ quarks is denoted by
$\Delta$. 
For instance, the $u$ quark density in the 2SC phase is given by three
contributions: unpaired blue quarks yield a term $\mu_u^3/(3\pi^2)$; paired
red and green quarks yield two terms, one contribution from a Fermi
sphere filled up to the average chemical potential $\nu$ plus
condensation energy proportional to $\Delta^2$.
In the last line we give $B^2/C$ to lowest order in $m_s$ and $\Delta$, which determines the
maximum bulk viscosity.
}
\label{tableBC}
\end{table}

\section{Derivation of the rate for the process $u+s\leftrightarrow u+d$}
\label{deriverate}

In this section we derive the rate $\Gamma_d$, see Eq.\ (\ref{p2}), as a function of $\d\m$, i.e., for
a system in chemical non-equilibrium. In equilibrium, $\d\m=0$ and $\Gamma_d=0$, since in this case
the processes $u+s\to u+d$ and $u+d\to u+s$ have the same rate and thus the net rate $\Gamma_d$ 
vanishes. We shall assume $\d\m=\mu_s-\mu_d>0$, such that we expect a net production of $d$ quarks,
$\Gamma_d>0$. The main result of this section is given by Eqs.\ (\ref{collision}), (\ref{afterang1}), 
and (\ref{afterang2}).

\subsection{Formalism}
\label{formalism}

\begin{figure}[ht]
\begin{center}
\includegraphics[width=0.7\textwidth]{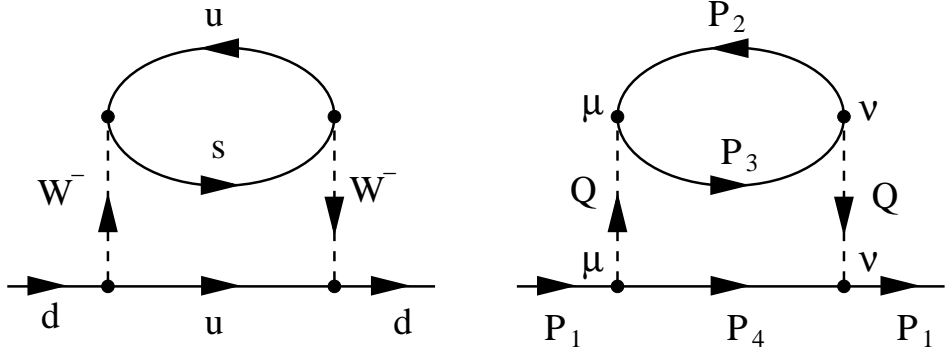}
%\vspace{0.5cm}
\caption{Contribution to the $d$-quark self-energy, necessary to compute the rate for the process
$u+s\leftrightarrow u+d$. The right diagram 
shows our notation of the corresponding momenta and Lorentz indices.}
\label{figself}
\end{center}
\end{figure}

We start from the kinetic equation which is derived from the Kadanoff-Baym equation \cite{KB} after 
applying a gradient expansion. For similar calculations within this formalism see  
Refs.~\cite{Jaikumar:2005hy,Schmitt:2005wg,friman,sedrakian}. 
Close to equilibrium and employing the closed-time-path formalism we may use the kinetic equation
\be \label{kinetic}
i\frac{\partial}{\partial t}{\rm Tr}[\g_0S^<(P_1)] = -{\rm Tr}[S^>(P_1)\Sigma^<(P_1) - 
\Sigma^>(P_1)S^<(P_1)] \, .
\ee
Here and in the following, four-momenta are denoted by $P\equiv(p_0,{\bf p})$. The quantities 
$S^{<,>}$ and $\Sigma^{<,>}$ are the fermion propagators and self-energies, respectively.
More precisely, they represent the $d$-quark propagator and the contribution to
the $d$-quark self-energy originating from the (left) diagram shown in Fig.~\ref{figself}. Then, 
the left-hand side of Eq.\ (\ref{kinetic}), integrated over $P_1$, basically is another way of
writing $\G_d$, while the right-hand side yields the collision integral for the process 
$u+s\leftrightarrow u+d$. Hence, in the following, we shall compute the rate $\G_d$ by evaluating
the right-hand side of Eq.\ (\ref{kinetic}). To this end, we first define $S^{<,>}$ and $\Sigma^{<,>}$
for the 2SC phase (remainder of this subsection and Sec.~\ref{defineprops}) and then insert these 
definitions back into the right-hand side of Eq.\ (\ref{kinetic}) to obtain the collision 
integral (Secs.~\ref{wbosontensor} and \ref{collisionintegral}).

Since we are describing a superconducting system, the fermion propagators are matrices in Nambu-Gorkov space,
\be \label{propnambu}
S^{<,>}(P) = \left(\begin{array}{cc} G_+^{<,>}(P) & F_-^{<,>}(P) \\ F_+^{<,>}(P) & G_-^{<,>}(P) 
\end{array}\right)\,.
\ee
The explicit form of the ``normal'' and ``anomalous'' propagators $G_\pm^{<,>}$ and $F_\pm^{<,>}$ 
shall be given in the next subsection
for the 2SC phase. The relevant part of the $d$-quark self-energy is given by the diagram in 
Fig.~\ref{figself} (the right diagram in this figure is merely a reminder of our notation of the momenta), 
\be \label{dquarkself}
\Sigma^{<,>}(P_1) = \frac{i}{M_W^4}
\int\frac{d^4P_4}{(2\pi)^4}\,\G_{ud,-}^\mu \,S^{<,>}(P_4)\,\G_{ud,+}^\nu\,
\Pi_{\mu\nu}^{>,<}(P_1-P_4) \, ,
\ee
where $M_W$ is the $W$-boson mass and $\G_{ud,\pm}^\mu$ is the vertex of the subprocess 
$d\leftrightarrow u + W^-$. The respective contribution to the $W$-boson polarization tensors are given by
\be \label{polardef}
\Pi_{\mu\nu}^{<,>}(Q) = -i \int\frac{d^4P_2}{(2\pi)^4}{\rm Tr}[\G_{us,+}^\mu \,S^{>,<}(P_2+Q)\,\G_{us,-}^\nu\,
S^{<,>}(P_2)] \, ,
\ee
where $\G_{us,\pm}^\mu$ is the vertex of the subprocess $u + W^-\leftrightarrow s$.
In Nambu-Gorkov space, the vertices are
\be
\G_{ud/us,\pm}^\mu = \frac{e\,V_{ud/us}}{2\sqrt{2}\sin\theta_W}\left(\begin{array}{cc} 
\g^\mu(1-\g^5)\,\tau_{ud/us,\pm} & 0 \\ 
0 &  -\g^\mu(1+\g^5)\,\tau_{ud/us,\mp}\end{array}\right) \, ,
\ee
with the entries of the CKM matrix $V_{ud}$ and $V_{us}$ and the weak mixing angle $\theta_W$, and 
\be
\tau_{ud,+} \equiv \left(\begin{array}{ccc}   0 & 1 &0 \\ 0 &0&0 \\ 0&0&0 \end{array}\right) \, , 
\quad \tau_{ud,-} \equiv \left(\begin{array}{ccc}   0 & 0 &0 \\ 1 &0&0 \\ 0&0&0 \end{array}\right) \, , 
\quad \tau_{us,+} \equiv \left(\begin{array}{ccc}   0 & 0 &1 \\ 0 &0&0 \\ 0&0&0 \end{array}\right) \, , 
\quad \tau_{us,-} \equiv \left(\begin{array}{ccc}   0 & 0 &0 \\ 0 &0&0 \\ 1&0&0 \end{array}\right) 
\ee
are matrices in flavor space (with indices ordered $u,d,s$).

\subsection{Quark propagators in the 2SC phase}
\label{defineprops}

We now specify the normal and anomalous propagators $G_\pm^{<,>}$ and $F_\pm^{<,>}$. We shall consider 
the simplest situation of massless quarks. In this case, the propagators in the 2SC phase are 
\be \label{2SCpropagators}
G_\pm^{<,>}(K) = \sum_{r=1}^3 {\cal P}_r G_{\pm,r}^{<,>}(K)\,\g^0\Lambda_{\bf k}^\mp \, , \quad
F_\pm^{<,>}(K) = J_3 I_3 F_{\pm,1}^{<,>}(K)\,\g^5\Lambda_{\bf k}^\mp \, . 
\ee
The index $r$ labels the different quasiparticle excitations, which are explained below and shown in Table 
\ref{tablemodes}. The quantities 
$\Lambda_{\bf k}^\mp\equiv (1\mp\g^0\vg\cdot\uk)/2$, with $\uk\equiv {\bf k}/k$, 
are energy projectors, $(I_i)_{jk} = (J_i)_{jk} = -i\e_{ijk}$ ($i,j,k\le 1,2,3$) are matrices in 
flavor and color space, respectively, and 
\begin{subequations}\label{greaterlesser}
\bea
G_{\pm,r}^>(K) &\equiv& -2\pi i\, \left\{B_{k,r}^\pm \, f(\ve_{k,r})\, \d(k_0\pm\mu_r-\ve_{k,r})
+B_{k,r}^\mp \, [1-f(\ve_{k,r})]\, \d(k_0\pm\mu_r+\ve_{k,r})\right\} \, , \\
G_{\pm,r}^<(K) &\equiv& -2\pi i\, \left\{B_{k,r}^\pm \, [1-f(\ve_{k,r})]\, \d(k_0\pm\mu_r-\ve_{k,r})
+B_{k,r}^\mp\, f(\ve_{k,r})\,\d(k_0\pm\mu_r+\ve_{k,r})\right\} \, ,\\
F_{\pm,r}^>(K) &\equiv& 2\pi i\,\frac{\Delta}{2\ve_{k,r}}\Big\{f(\ve_{k,r})\,\d(k_0\mp\mu_r-\ve_{k,r})
-[1-f(\ve_{k,r})]\,\d(k_0\mp\mu_r+\ve_{k,r})\Big\} \, ,\\
F_{\pm,r}^<(K) &\equiv& 2\pi i\,\frac{\Delta}{2\ve_{k,r}}\Big\{[1-f(\ve_{k,r})]\,\d(k_0\pm\mu_r-\ve_{k,r})
-f(\ve_{k,r})\,\d(k_0\pm\mu_r+\ve_{k,r})\Big\} \, .
\eea
\end{subequations}
%\begin{subequations}\label{greaterlesser}
%\bea
%G_{\pm,r}^>(K) &\equiv& i\pi \frac{k_0\mp k}{\ve_{k,r}}\,\left[f(\ve_{k,r})\,
%\delta(k_0\mp\mu_r-\ve_{k,r}) - f(-\ve_{k,r})\, \delta(k_0\mp\mu_r+\ve_{k,r})\right] \, ,\\
%G_{\pm,r}^<(K) &\equiv& -i\pi \frac{k_0\mp k}{\ve_{k,r}}\,\left[f(\ve_{k,r})\,
%\delta(k_0\mp\mu_r+\ve_{k,r}) - f(-\ve_{k,r})\, \delta(k_0\mp\mu_r-\ve_{k,r})\right] \, ,\\
%F_{\pm,r}^>(K) &\equiv& i\pi \frac{\Delta}{\ve_{k,r}}\,\left[f(\ve_{k,r})\,
%\delta(k_0-\ve_{k,r}) - f(-\ve_{k,r})\, \delta(k_0+\ve_{k,r})\right]\, ,\\
%F_{\pm,r}^<(K) &\equiv& - i\pi \frac{\Delta}{\ve_{k,r}}\,\left[f(\ve_{k,r})\,
%\delta(k_0+\ve_{k,r}) - f(-\ve_{k,r})\, \delta(k_0-\ve_{k,r})\right] \, ,
%\eea
%\end{subequations}
Here, we have introduced the Bogoliubov coefficients
\be \label{bogoliubov}
B_{k,r}^\pm\equiv \frac{\ve_{k,r}\pm (\mu_r-k)}{2\ve_{k,r}} \, .
\ee
Moreover, $f$ is the fermion distribution function. For the right-hand side of the kinetic equation, we
shall approximate $f$ by its equilibrium form,
\be \label{Fermi}
f(\ve)\equiv \frac{1}{e^{\ve/T}+1} \, .
\ee
We have neglected the antiparticle contribution to the propagators.
For simplicity, we assume 
$\mu_u=\mu_d\equiv\mu$. This is somewhat unrealistic since it violates overall electric charge neutrality.
However, a more realistic scenario $\mu_u\neq \mu_d$ would lead to some additional 
(notational) complication and does not alter our results qualitatively. 
With this simplification, we have 3 different quasiparticle dispersion relations, $\ve_{k,r}$,
with $r=1,2,3$. They are shown in Table \ref{tablemodes} with $\Delta$ being the energy gap induced by 
Cooper pairing of $u$ and $d$ quarks. The condensate picks a certain direction in color
and flavor space, here $J_3$ and $I_3$, respectively, leaving one color of $u$ and $d$ quarks as 
well as all $s$ quarks unpaired.  
Formally, each of the three modes lives in a subspace of the 9-dimensional color-flavor space, which is
given by the corresponding projectors ${\cal P}_i$ occurring in the normal propagators $G_\pm^{<,>}(K)$.
The projectors are shown in the last column of Table \ref{tablemodes}.
The dimensions of the subspaces are 4, 2, 3, respectively. 
In the anomalous propagators, only the gapped modes, $r=1$, contribute. 
\begin{table}  
\begin{tabular}[t]{|c||c|c|c|}
\hline
$\;\;$ $r$$\;\;$ & quarks & dispersion & projector \\ \hline\hline
1  & $\;\;$$ru$, $rd$, $gu$, $gd$$\;\;$ &$\;\;$ $\ve_{k,1}=\sqrt{(k-\mu)^2+\Delta^2}$$\;\;$ 
& ${\cal P}_1=I_3^2J_3^2$ \\ \hline
2 & $bu$, $bd$ & $\ve_{k,2}=|k-\mu|$ & $\;\;$${\cal P}_2=I_3^2(1-J_3^2)$$\;\;$\\ \hline
3 & $rs$, $gs$, $bs$ & $\ve_{k,3} = |k-\mu_s|$&${\cal P}_3=1-I_3^2$
\\ \hline
\end{tabular}
\caption{Quasiparticle modes in the 2SC phase for ultra-relativistic quarks and $\mu_u=\mu_d\equiv\mu$.}
\label{tablemodes}
\end{table}

\subsection{$W$-boson polarization tensor}  
\label{wbosontensor}

Next, we compute the contribution to the $W$-boson polarization tensor that occurs in the 
$d$-quark self-energy, Eq.\ (\ref{dquarkself}). From its definition (\ref{polardef}) 
we obtain after inserting Eq.\ (\ref{propnambu}) and performing the trace over Nambu-Gorkov space
\bea
\Pi_{\mu\nu}^{<,>}(Q) &=& -i \frac{e^2V_{us}^2}{8\sin^2\theta_W}\int\frac{d^4P_2}{(2\pi)^4}\Big\{
{\rm Tr}[\g^\mu(1-\g^5)\,\tau_{us,+}\,G_+^{>,<}(P_2+Q)\,\g^\nu(1-\g^5)\,\tau_{us,-}\,G_+^{<,>}(P_2)] \non
&&+\;{\rm Tr}[\g^\mu(1+\g^5)\,\tau_{us,-}\,G_-^{>,<}(P_2+Q)\,\g^\nu(1+\g^5)\,\tau_{us,+}\,G_-^{<,>}(P_2)] \non
&&-\;{\rm Tr}[\g^\mu(1-\g^5)\,\tau_{us,+}\,F_-^{>,<}(P_2+Q)\,\g^\nu(1+\g^5)\,\tau_{us,+}\,F_+^{<,>}(P_2)] \non
&&-\;{\rm Tr}[\g^\mu(1+\g^5)\,\tau_{us,-}\,F_+^{>,<}(P_2+Q)\,\g^\nu(1-\g^5)\,\tau_{us,-}\,F_-^{<,>}(P_2)] 
\Big\}
\, .\label{afternambu}
\eea
We assume without explicit proof that the second (fourth) term yields the same contribution as the first 
(third) term because this is just the charge-conjugate counterpart. Next we use Eq.\ (\ref{2SCpropagators}) 
and perform the trace over color and flavor space. We make use of
\begin{subequations}
\bea
\tau_{us,+}{\cal P}_1 &=& 0 \,\, , \qquad \tau_{us,-}{\cal P}_1=\tau_{us,-}J_3^2 \, ,\\
\tau_{us,+}{\cal P}_2 &=& 0 \,\, , \qquad \tau_{us,-}{\cal P}_2=\tau_{us,-}(1-J_3^2) \, ,\\
\tau_{us,+}{\cal P}_3 &=& \tau_{us,+} \,\, , \qquad \tau_{us,-}{\cal P}_3=0 \, ,\\
\tau_{us,\pm}I_3\tau_{us,\pm}I_3 &=& 0 \, . \label{anomvanish}
\eea
\end{subequations}
Because of Eq.\ (\ref{anomvanish}), the anomalous 
propagators yield no contribution. Then, with $P_3= P_2+Q$ and denoting  
\be \label{diracT}
{\cal T}^{\mu\nu}(\up,\uk)\equiv {\rm Tr}[\g^\mu(1-\g^5)\g^0\Lambda_{\bf p}^-\g^\nu(1-\g^5)\g^0\Lambda_{\bf k}^-]\, ,
\ee
we obtain 
\be
\Pi_{\mu\nu}^{<,>}(Q) = -2i \frac{e^2V_{us}^2}{8\sin^2\theta_W}\int\frac{d^4P_2}{(2\pi)^4}
{\cal T}^{\mu\nu}(\up_3,\up_2) \left[2G_{+,3}^{>,<}(P_3)G_{+,1}^{<,>}(P_2) + 
G_{+,3}^{>,<}(P_3)G_{+,2}^{<,>}(P_2)\right]\,
 \, .
\ee
Inserting the propagators from Eqs.\ (\ref{greaterlesser}) into this expression and using 
$1-f(\ve)=f(-\ve)$ for the equilibrium function (\ref{Fermi}) yields
%\begin{subequations}
%\bea
%\int_{-\infty}^{\infty}\frac{dp_{20}}{2\pi}G_{+,3}^>(P_3)G_{+,r}^<(P_2) &=& 
%-2\pi\sum_{e_2e_3}B_{p_3,3}^{e_3}B_{p_2,r}^{e_2}
%\, f(e_3\ve_{p_3,3})f(-e_2\ve_{p_2,r})\,\d(q_0-e_3\ve_{p_3,3}+e_2\ve_{p_2,r}-\d\m) \, ,\\
%\int_{-\infty}^{\infty}\frac{dp_{20}}{2\pi}G_{+,3}^<(P_3)G_{+,r}^>(P_2) &=& 
%-2\pi\sum_{e_2e_3}B_{p_3,3}^{e_3}B_{p_2,r}^{e_2}
%\, f(-e_3\ve_{p_3,3})f(e_2\ve_{p_2,r})\,\d(q_0-e_3\ve_{p_3,3}+e_2\ve_{p_2,r}-\d\m) \, ,
%\eea
%\end{subequations}
%where the sum runs over $e_2,e_3=\pm 1$.
%Consequently, the polarization tensors become
\bea \label{Pi}
\Pi_{\mu\nu}^{<,>}(Q)&=& 4\pi i \frac{e^2V_{us}^2}{8\sin^2\theta_W}\sum_{e_2e_3}\int\frac{d^3{\bf p}_2}{(2\pi)^3}
{\cal T}^{\mu\nu}(\up_3,\up_2) \non 
&&\times\;\Big\{2B_{p_3,3}^{e_3}B_{p_2,1}^{e_2}
\, f(\pm e_3\ve_{p_3,3})f(\mp e_2\ve_{p_2,1})\,\d(q_0-e_3\ve_{p_3,3}+e_2\ve_{p_2,1}+\d\mu) \non
&&+\; 
B_{p_3,3}^{e_3}B_{p_2,2}^{e_2}
\, f(\pm e_3\ve_{p_3,3})f(\mp e_2\ve_{p_2,2})\,\d(q_0-e_3\ve_{p_3,3}+e_2\ve_{p_2,2}+\d\m)\Big\} \, ,
\eea
where the sum runs over $e_2,e_3=\pm 1$, and the upper (lower) sign corresponds to $\Pi^<$ ($\Pi^>$). 
Remember that the quark loop in the polarization tensor contains a $u$ and an $s$ quark. Therefore, 
both contributions in curly brackets contain the third mode, $r=3$, describing an (ungapped) $s$ quark.
For the $u$ quark, there are two options. The first term in curly brackets describes a gapped $u$ quark,
$r=1$. The factor 2 in front of this term originates from color degeneracy: Both 
red and green $u$ quarks are gapped. The second term in curly brackets describes an ungapped blue
$u$ quark.

\subsection{Collision integral}
\label{collisionintegral}

Upon inserting Eq.\ (\ref{dquarkself}) into Eq.\ (\ref{kinetic}), the kinetic equation becomes
\bea
i\frac{\partial}{\partial t}{\rm Tr}[\g_0S^<(P_1)] &=& -\frac{i}{M_W^4}\int\frac{d^4P_4}{(2\pi)^4}\,
{\rm Tr}[S^>(P_1)\G_{ud,-}^\mu S^<(P_4) \G_{ud,+}^\nu \Pi^>_{\m\n}(P_1-P_4) \non
&&\hspace{2cm}-\, \G_{ud,-}^\mu S^>(P_4)\G_{ud,+}^\nu S^<(P_1) \Pi^<_{\m\n}(P_1-P_4))] \, .
\eea
Hence, in order to evaluate the right-hand side of this equation, we need
the traces
\bea
{\rm Tr}[S^>(P_1)\G_{ud,-}^\mu S^<(P_4)\G_{ud,+}^\nu] &=& \frac{e^2V_{ud}^2}{8\sin^2\theta_W}
\Big\{ {\rm Tr}[G_+^>(P_1)\g^\m(1-\g^5)\tau_{ud,-}G_+^<
(P_4)\g^\nu(1-\g^5)\tau_{ud,+}] \non
&&+\;{\rm Tr}[G_-^>(P_1)\g^\m(1+\g^5)\tau_{ud,+}G_-^<(P_4)\g^\nu(1+\g^5)\tau_{ud,-}]\non
&&-\;{\rm Tr}[F_-^>(P_1)\g^\m(1+\g^5)\tau_{ud,+}F_+^<(P_4)\g^\nu(1-\g^5)\tau_{ud,+}]\non
&&-\;{\rm Tr}[F_+^>(P_1)\g^\m(1-\g^5)\tau_{ud,-}F_-^<(P_4)\g^\nu(1+\g^5)\tau_{ud,-}]\Big\} \, ,
\label{afternambu21}
\eea
and
\bea
{\rm Tr}[\G_{ud,-}^\mu S^>(P_4) \G_{ud,+}^\nu S^<(P_1)] &=& \frac{e^2V_{ud}^2}{8\sin^2\theta_W}
\Big\{{\rm Tr}[\g^\m(1-\g^5)\tau_{ud,-}G_+^>(P_4)
\g^\nu(1-\g^5)\tau_{ud,+} G_+^<(P_1)] \non
&&+\;{\rm Tr}[\g^\m(1+\g^5)\tau_{ud,+}G_-^>(P_4)\g^\nu(1+\g^5)\tau_{ud,-} G_-^<(P_1)] \non
&&-\;{\rm Tr}[\g^\m(1-\g^5)\tau_{ud,-}F_-^>(P_4)\g^\nu(1+\g^5)\tau_{ud,-} F_+^<(P_1)] \non
&&-\;{\rm Tr}[\g^\m(1+\g^5)\tau_{ud,+}F_+^>(P_4)\g^\nu(1-\g^5)\tau_{ud,+} F_-^<(P_1)]\Big\} \, .
\label{afternambu22}
\eea
Again, in both expressions, the first (third) and second (fourth) traces yield identical contributions.
For the traces over color and flavor space we 
make use of
\begin{subequations}
\bea
\tau_{ud,\pm}{\cal P}_1 &=& \tau_{ud,\pm} J_3^2 \, ,\\
\tau_{ud,\pm}{\cal P}_2 &=& \tau_{ud,\pm} (1-J_3^2) \, ,\\
\tau_{ud,\pm}{\cal P}_3 &=& 0 \, ,\\
\tau_{ud,+}I_3\tau_{ud,+}I_3 &=& I_1^2-1 \\
\tau_{ud,-}I_3\tau_{ud,-}I_3 &=& I_2^2-1 \, .
\eea
\end{subequations}
This shows that, in contrast to the $W$-boson polarization tensor,
the traces containing anomalous propagators do not vanish. We obtain
\bea \label{rhs1}
{\rm Tr}[S^>(P_1)\Sigma^<(P_1)]&=& i\frac{e^2V_{ud}^2}{4M^4_W\sin^2\theta_W}
\int\frac{d^4P_4}{(2\pi)^4}\Big\{\left[2G_{+,1}^>(P_1)G_{+,1}^<(P_4) + 
G_{+,2}^>(P_1)G_{+,2}^<(P_4)\right]{\cal T}^{\m\n}(\up_4,\up_1) \non
&&+\; 2F_{-,1}^>(P_1)F_{+,1}^<(P_4)
\,{\cal U}^{\n\m}(\up_1,\up_4)\Big\}\,\Pi_{\m\n}^>(P_1-P_4) \, ,
\eea
and
\bea \label{rhs2}
{\rm Tr}[\Sigma^>(P_1)S^<(P_1)]&=& i\frac{e^2V_{ud}^2}{4M_W^4\sin^2\theta_W}
\int\frac{d^4P_4}{(2\pi)^4}\Big\{\left[2G_{+,1}^>(P_4)G_{+,1}^<(P_1) + 
G_{+,2}^>(P_4)G_{+,2}^<(P_1)\right]{\cal T}^{\m\n}(\up_4,\up_1) \non
&&+\; 2F_{-,1}^>(P_4)F_{+,1}^<(P_1)
\,{\cal U}^{\m\n}(\up_4,\up_1)\Big\}\,\Pi_{\m\n}^<(P_1-P_4) \, ,
\eea
where we defined the trace over Dirac space,
\be \label{diracU}
{\cal U}^{\mu\nu}(\up,\uk)\equiv {\rm Tr}[\g^\mu(1-\g^5)\g^5\Lambda_{{\bf p}}^+\g^\nu
(1+\g^5)\g^5\Lambda_{{\bf k}}^-] \, .
\ee
Again, the different contributions in curly brackets in Eqs.\ (\ref{rhs1}) and (\ref{rhs2}) are easy to 
understand. Remember that these expressions describe the subprocess $d\leftrightarrow W^- + u$.
Therefore, one contribution comes from the gapped $u$ and $d$ quarks, $r=1$. This mode also 
yields a contribution from the anomalous propagators. It describes red and green quarks, hence the 
prefactor 2. The second contribution comes from the ungapped mode, $r=2$, which describes 
blue quarks and which, of course, yields no 
anomalous contribution. Due to color conservation of the weak interaction, no mixing of the two 
modes is possible.

Next, we perform the integration over $p_{40}$. Moreover, we shall integrate the kinetic equation 
over $d$-quark energies $p_{10}$ and momenta ${\bf p}_1$, because we are interested in the total rate, not in 
the rate as a function of momentum. Therefore, we need the results (derived by inserting the propagators
from Eqs.\ (\ref{greaterlesser}) into Eqs.\ (\ref{rhs1}) and (\ref{rhs2}))
\bea 
\int_{-\infty}^{\infty}\frac{dp_{10}}{2\pi}{\rm Tr}[S^>(P_1)\Sigma^<(P_1)]
&=&-i\frac{e^2V_{ud}^2}{4M_W^4\sin^2\theta_W}\sum_{e_1e_4}\int\frac{d^3{\bf p}_4}{(2\pi)^3}\Bigg\{2\left[B_{p_1,1}^{e_1}B_{p_4,1}^{e_4}\,
{\cal T}^{\mu\nu}(\up_4,\up_1) + \frac{e_1e_4\Delta^2}{4\ve_{p_1,1}\ve_{p_4,1}}{\cal U}^{\n\m}(\up_1,\up_4)
\right]\non
&& \hspace{-1cm}\times\;f(e_1\ve_{p_1,1})f(-e_4\ve_{p_4,1})\,
\Pi_{\mu\nu}^>(e_1\ve_{p_1,1}-e_4\ve_{p_4,1},{\bf p}_1-{\bf p}_4)\non
&&\hspace{-1cm}+\,B_{p_1,2}^{e_1}B_{p_4,2}^{e_4}\,
{\cal T}^{\mu\nu}(\up_4,\up_1)\,f(e_1\ve_{p_1,2})f(-e_4\ve_{p_4,2})\,
\Pi_{\mu\nu}^>(e_1\ve_{p_1,2}-e_4\ve_{p_4,2},{\bf p}_1-{\bf p}_4)\Bigg\} \,,
\label{trSSigma}
\eea
and
\bea
\int_{-\infty}^{\infty}\frac{dp_{10}}{2\pi}{\rm Tr}[\Sigma^>(P_1)S^<(P_1)]
&=&-i\frac{e^2V_{ud}^2}{4M_W^4\sin^2\theta_W}\sum_{e_1e_4}\int\frac{d^3{\bf p}_4}{(2\pi)^3}
\Bigg\{2\left[B_{p_1,1}^{e_1}B_{p_4,1}^{e_4}\,
{\cal T}^{\mu\nu}(\up_4,\up_1) + \frac{e_1e_4\Delta^2}{4\ve_{p_1,1}\ve_{p_4,1}}{\cal U}^{\mu\nu}(\up_4,\up_1)
\right]\non
&&\hspace{-1cm}\times\;f(-e_1\ve_{p_1,1})f(e_4\ve_{p_4,1})\,
\Pi_{\mu\nu}^<(e_1\ve_{p_1,1}-e_4\ve_{p_4,1},{\bf p}_1-{\bf p}_4)\non
&& \hspace{-1cm}+\,B_{p_1,2}^{e_1}B_{p_4,2}^{e_4}\,
{\cal T}^{\mu\nu}(\up_4,\up_1)\,f(-e_1\ve_{p_1,2})f(e_4\ve_{p_4,2})\,
\Pi_{\mu\nu}^<(e_1\ve_{p_1,2}-e_4\ve_{p_4,2},{\bf p}_1-{\bf p}_4)\Bigg\} \, .
\label{trSigmaS} 
\eea
%where
%\be
%Q^r_{e_1e_4}\equiv (-e_1\ve_{p_1,r}+e_4\ve_{p_4,r},{\bf p}_1-{\bf p}_4) \,\, , \qquad r=1,2 \,.
%\ee
Then, in order to obtain the right-hand side of Eq.\ (\ref{kinetic}) (integrated over $P_1$), we 
have to insert Eq.\ (\ref{Pi}) into Eqs.\ (\ref{trSSigma}) and (\ref{trSigmaS}) and the result
into Eq.\ (\ref{kinetic}). One needs the explicit results for the Dirac traces and contractions 
of the form ${\cal T}^{\m\n}{\cal T}_{\m\n}$ and ${\cal U}^{\m\n}{\cal T}_{\m\n}$. 
They are given in Appendix \ref{appdirac}. As shown in this appendix, the contractions of the
form ${\cal U}^{\m\n}{\cal T}_{\m\n}$, i.e., the ones originating from the contribution of anomalous 
propagators, are complex. However, after the angular integration in the collision integral, the
imaginary parts vanish, see Appendix \ref{appcomplex}. Therefore, we shall continue with
only the real parts of these contractions.

As mentioned above, the left-hand side of the kinetic equation (\ref{kinetic}) stands for $\Gamma_d$, the 
change of the total number of $d$ quarks per time and volume (as given by the right-hand side). 
In order to define $\Gamma_d$ properly and extract the correct numerical factor, we integrate the 
left-hand side over $P_1$ and project onto the $d$-flavor subspace. The latter is done by inserting 
the projector ${\cal Q}_d\equiv {\rm diag}(0,1,0)$. Moreover, the change in density is obtained upon 
multiplying by the matrix $\tau_3\equiv{\rm diag}(1,-1)$ in Nambu-Gorkov space (see
for instance Ref.\ \cite{vollhardt}). Then, using the propagators from 
Eqs.\ (\ref{greaterlesser}), we obtain    
\bea \label{leftside}
i\frac{\partial}{\partial t} \int\frac{d^4P_1}{(2\pi)^4}{\rm Tr}[\g_0{\cal Q}_d \tau_3 S^<(P_1)]
&=& -4\frac{\partial}{\partial t}\int\frac{d^3{\bf p}_1}{(2\pi)^3}\left\{2B^-_{p_1,1}f(\ve_{p_1,1})
+2B^+_{p_1,1}[1-f(\ve_{p_1,1})] \right. \non
&&\left.\hspace{2.2cm}+\,B_{p_1,2}^- f(\ve_{p_1,2}) + B_{p_1,2}^+ [1-f(\ve_{p_1,2})]\right\} \non
&\equiv& -4\frac{\partial}{\partial t} \int\frac{d^3{\bf p}_1}{(2\pi)^3}f_d(t,{\bf p}_1)\equiv -4\G_d \, ,
\eea
where we have used $B_{p_1,r}^+ + B_{p_1,r}^-=1$. 
Eq.\ (\ref{leftside}) is the microscopic definition of $\G_d$ which has been used implicitly in 
Sec.~\ref{definebulk}.
The factor 4 arises after performing the trace; it originates from the Nambu-Gorkov doubling and spin
degeneracy. 

%\old{The left-hand side of the kinetic equation has also to be integrated over $P_1$. Furthermore,
%we have to project onto the $d$-flavor subspace. This is done by inserting the projector
%${\cal Q}_d\equiv {\rm diag}(0,1,0)$. We obtain 
%\bea
%i\frac{\partial}{\partial t} \int\frac{dP_1}{(2\pi)^4}{\rm Tr}[\g_0{\cal Q}_d S^<(P_1)]
%&=& -4\frac{\partial}{\partial t}\int\frac{d^3{\bf p}_1}{(2\pi)^3}\left\{2B^+_{p_1,1}f(\ve_{p_1,1})
%+2B^-_{p_1,1}[1-f(\ve_{p_1,1})] \right. \non
%&&\left.\hspace{2.2cm}+\,B_{p_1,2}^+ f(\ve_{p_1,2}) + B_{p_1,2}^- [1-f(\ve_{p_1,2})]\right\} \non
%&\equiv& -4\frac{\partial}{\partial t} \int\frac{d^3{\bf p}_1}{(2\pi)^3}f_d(t,{\bf p}_1)\equiv -4\G_d \, ,
%\eea
%where $f_d(t,{\bf p}_1)$ is the (non-equilibrium) $d$-flavor distribution function. Hence, 
%the rate $\G_d$ is the change of the total number of $d$ quarks per time and volume.
%This is the microscopic definition of $\G_d$ which implicitly has been used in Sec.~\ref{definebulk}.
%The factor 4 arises after performing the trace; it originates from the Nambu-Gorkov doubling and spin
%degeneracy.}

 Putting the results for the left-hand and right-hand sides 
together and dividing both sides by 4 yields 
\be \label{collision}
\G_d=
%\G_d(\mu,\d\m,T) = 
4\G_d^{1131}  + 2 \G_d^{1231}+2\G_d^{2132} + \G_d^{2232}+ 
4 \widetilde{\G}_d^{1131}+ 2\widetilde{\G}_d^{1231}  
 \, . 
\ee
We shall explain the physical meaning of the several terms in the next subsection. Their explicit form is
\bea \label{gammad}
\G_d^{r_1r_2r_3r_4}&\equiv& 128\pi^4 G_F^2V^2_{ud}V^2_{us}\sum_{e_1e_2e_3e_4} 
\int\frac{d^3{\bf p}_1}{(2\pi)^3}\int\frac{d^3{\bf p}_2}{(2\pi)^3}\int\frac{d^3{\bf p}_3}{(2\pi)^3}
\int\frac{d^3{\bf p}_4}{(2\pi)^3}\,\delta({\bf p}_1+{\bf p}_2-{\bf p}_3-{\bf p}_4) \non
&&\times\,(1-\up_1\cdot\up_2)(1-\up_3\cdot\up_4)\,B^{e_1}_1B^{e_2}_2B^{e_3}_3B^{e_4}_4 
\delta(e_1\ve_1+e_2\ve_2-e_3\ve_3-e_4\ve_4+\d\m) \non
&&\times\,\left[f(e_1\ve_1)f(e_2\ve_2)f(-e_3\ve_3)f(-e_4\ve_4)
-f(-e_1\ve_1)f(-e_2\ve_2)f(e_3\ve_3)f(e_4\ve_4)\right] \, ,
\eea
and 
\bea \label{tildegammad}
\widetilde{\G}_d^{r_1r_2r_3r_4}&\equiv& 64\pi^4 G_F^2V^2_{ud}V^2_{us}\sum_{e_1e_2e_3e_4} 
\int\frac{d^3{\bf p}_1}{(2\pi)^3}\int\frac{d^3{\bf p}_2}{(2\pi)^3}\int\frac{d^3{\bf p}_3}{(2\pi)^3}
\int\frac{d^3{\bf p}_4}{(2\pi)^3}\,\delta({\bf p}_1+{\bf p}_2-{\bf p}_3-{\bf p}_4) \non
&&\times\,\left[(1-\up_4\cdot\up_3)(1-\up_1\cdot\up_2)+(1+\up_1\cdot\up_3)(1+\up_4\cdot\up_2)
-(1+\up_1\cdot\up_4)(1+\up_3\cdot\up_2)\right]\, \non
&&\times\,\frac{e_1\Delta}{2\ve_1}
B_2^{e_2}B_3^{e_3}\frac{e_4\Delta}{2\ve_4}
\delta(e_1\ve_1+e_2\ve_2-e_3\ve_3-e_4\ve_4+\d\m) \non
&&\times\,\left[f(e_1\ve_1)f(e_2\ve_2)f(-e_3\ve_3)f(-e_4\ve_4)
-f(-e_1\ve_1)f(-e_2\ve_2)f(e_3\ve_3)f(e_4\ve_4)\right] \, ,
\eea
with the Fermi coupling constant $G_F$.
We have introduced an additional integration over ${\bf p}_3$ and the corresponding $\d$-function for 
momentum conservation. This was implicitly present in the previous expressions. Moreover, we have abbreviated
the quasiparticle energies with momentum $p_i$ and excitation branch $r_i$ by 
$\ve_i\equiv \ve_{p_i,r_i}$ and the Bogoliubov coefficients by $B_i^{e_i}\equiv B_{p_i,r_i}^{e_i}$. 
We may now perform the angular integration.
This calculation is identical to the unpaired phase \cite{wadhwa}, see Appendix \ref{appangular}.  The results
are
\bea \label{afterang1}
\G_d^{r_1r_2r_3r_4}&\equiv& \frac{G_F^2V^2_{ud}V^2_{us}}{8\pi^5}\sum_{e_1e_2e_3e_4} 
\int_{p_1p_2p_3p_4} \,I(p_1,p_2,p_3,p_4)
\,B^{e_1}_1B^{e_2}_2B^{e_3}_3B^{e_4}_4 \, \delta(e_1\ve_1+e_2\ve_2-e_3\ve_3-e_4\ve_4+\d\m) \non
&&\times\,\left[f(e_1\ve_1)f(e_2\ve_2)f(-e_3\ve_3)f(-e_4\ve_4)
-f(-e_1\ve_1)f(-e_2\ve_2)f(e_3\ve_3)f(e_4\ve_4)\right] \, ,
\eea
and 
\bea \label{afterang2}
\widetilde{\G}_d^{r_1r_2r_3r_4}&\equiv& \frac{G_F^2V^2_{ud}V^2_{us}}{16\pi^5} \sum_{e_1e_2e_3e_4} 
\int_{p_1p_2p_3p_4} \,\tilde{I}(p_1,p_2,p_3,p_4)
\,\frac{e_1\Delta}{2\ve_1}
B_2^{e_2}B_3^{e_3}\frac{e_4\Delta}{2\ve_4} \,\delta(e_1\ve_1+e_2\ve_2-e_3\ve_3-e_4\ve_4+\d\m) \non
&&\times\,\left[f(e_1\ve_1)f(e_2\ve_2)f(-e_3\ve_3)f(-e_4\ve_4)
-f(-e_1\ve_1)f(-e_2\ve_2)f(e_3\ve_3)f(e_4\ve_4)\right] \, ,
\eea
where the functions $I$ and $\tilde{I}$ are given in Eqs.\ (\ref{IItilde}) and where we abbreviated
\be
\int_{p_1p_2p_3p_4} \equiv \int_0^\infty dp_1\int_0^\infty dp_2\int_0^\infty dp_3\int_0^\infty dp_4 \,.
\ee

\subsection{Subprocesses and unpaired phase limit}
\label{subprocesses}

\begin{figure}[ht]
\begin{center}
\includegraphics[width=\textwidth]{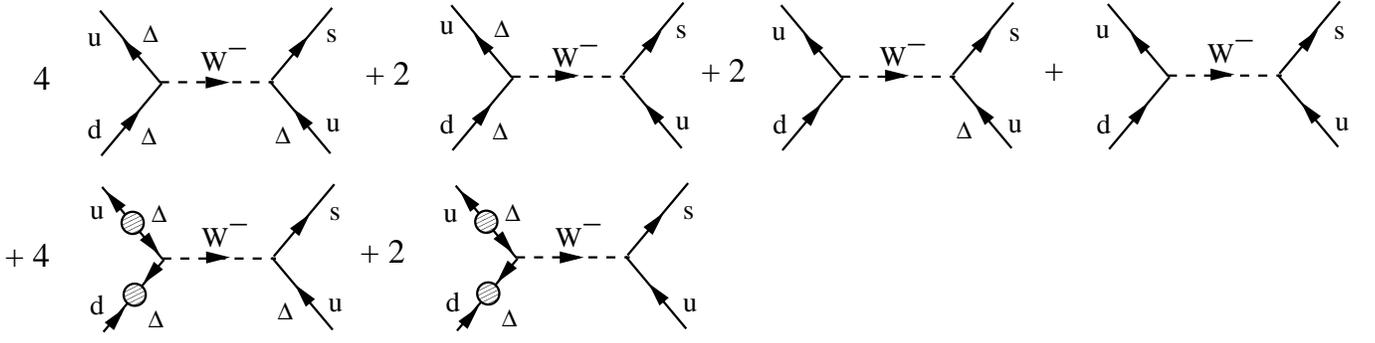}
%\vspace{0.5cm}
\caption{Contributions to the process $d+u\to u+s$. The four diagrams in the first row contain 
3, 2, 1, and 0 gapped fermions, respectively (a gapped fermion is marked by $\Delta$ at the respective 
lines). The combinations of quarks of the first two diagrams in the first row
allow for nonvanishing contributions of anomalous propagators. These are shown in the second row 
(a condensate insertion is represented by a hatched circle).}
\label{figdiagrams}
\end{center}
\end{figure}

The collision integral is the sum of six terms, see Eq.\ (\ref{collision}), which are 
shown diagrammatically in Fig.~\ref{figdiagrams} (in the same order as they appear in the equation).
The superscripts at $\G_d$ and $\widetilde{\G}_d$ in Eq.\ (\ref{collision}) correspond to the momenta 
in the order 
$p_1$, $p_2$, $p_3$, $p_4$, i.e., they describe the particles $d$, $u_1$, $s$, $u_2$, respectively,
see diagrams in Fig.~\ref{figself}. The combinatorics of the process is constrained by 
color conservation for the vertices $d\leftrightarrow u_2 + W^-$ and $u_1 + W^-\leftrightarrow s$.
Consequently, $d$ and $u_2$ as well as $s$ and $u_1$ must have the same color. Hence, from Table 
\ref{tablemodes} we conclude that $d$ and $u_2$ must be either both gapped ($r=1$) or both ungapped ($r=2$).
This can be seen in Fig.~\ref{figdiagrams}, where we have added the gap $\Delta$ at the fermion line
whenever the corresponding mode is gapped. It can also be seen in Eq.\ (\ref{collision}), where the 
pair of first and last superscripts is either (1,1) or (2,2). Furthermore, the second and third superscript
can be (1,3) or (2,3), describing a (red or green) gapped $u$ and a (red or green) $s$ or a (blue) 
ungapped $u$ and a (blue) $s$, respectively. The $s$ quark is ungapped for all colors. Combining these 
pairs and counting color degeneracies yields the six terms with respective prefactors in 
Eq.\ (\ref{collision}) and Fig.~\ref{figdiagrams}. In the cases where both $d$ and $u_2$ are gapped, there 
is an additional contribution from the anomalous propagators. This is denoted by $\widetilde{\Gamma}_d$ in 
Eq.\ (\ref{collision}) and by a condensate insertion in Fig.~\ref{figdiagrams}. From the diagrams
it is clear that there can be no condensate insertion for $u_1$. This would lead to a violation 
of electric charge conservation at the vertex $u_1 + W^-\leftrightarrow s$. 
Summarizing the result of the combinatorics, we have found that the process $d+u_1\leftrightarrow s+u_2$ 
in the 2SC phase can be considered as composed of several subprocesses, where zero, one, two, or 
three of the participating particles have a gap in their excitation spectrum.

\begin{figure*} [ht]
\begin{center}
\includegraphics[width=0.45\textwidth]{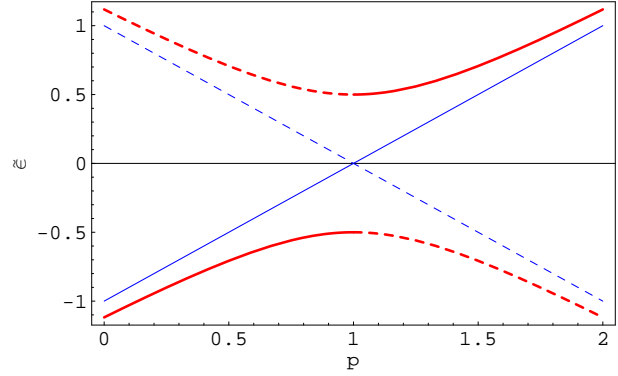}
\caption{(Color online)  Thick (red) lines: gapped modes. Thin (blue) lines: ungapped modes.  Solid lines:
$+\,\tilde{\ve}$. Dashed lines: $-\,\tilde{\ve}$. The quantities are given in arbitrary units,
$\mu=1$.
This figure illustrates the new variables introduced in Eq.\ (\ref{newbogol}), which are 
advantageous for the discussion of the various subprocesses, as explained 
below Eq.\ (\ref{normalphase}).}
\label{figparticlehole}
\end{center}
\end{figure*}

From Eqs.\ (\ref{afterang1}) and (\ref{afterang2}) we see that the process 
$d+u_1\leftrightarrow s+u_2$ not only contains several subprocesses due to the 2SC pairing pattern, 
also every single of these subprocesses is written as a sum over 16 terms, expressed as 
a sum over $e_1,\ldots, e_4$. To understand the physical meaning of this sum, it is instructive to rewrite 
it and recover the unpaired phase limit. The change of variables described in the following is illustrated in 
Fig.\ \ref{figparticlehole}. We may rewrite every single sum over $e_i$ by observing
that for any function $F$ we have
\be
\sum_{e_i}\int_0^\infty dp_i\, B_i^{e_i} F(e_i\ve_i) =
\sum_{e_i}\int_0^\infty dp_i\,\tilde{B}_i^{e_i} F(-e_i\tilde{\ve}_i) \, , 
\ee
where the new Bogoliubov coefficients and quasiparticle energies are defined as 
\be \label{newbogol}
\tilde{B}_i^{e_i} \equiv \frac{1}{2}\left(1+e_i\frac{p_i-\mu_i}{\tilde{\ve}_i}\right) \, ,
\qquad \tilde{\ve}_i\equiv {\rm sgn}(p_i-\mu_i)\ve_i \, .
\ee
Consequently, in Eqs.\ (\ref{afterang1}) and (\ref{afterang2}) we can replace $B_i^{e_i}$ by 
$\tilde{B}_i^{e_i}$ and $e_i\ve_i$ by $-e_i\tilde{\ve}_i$. Then, we obtain for the contribution from
the normal propagators,
\bea \label{tilde1}
\G_d^{r_1r_2r_3r_4}&\equiv& \frac{G_F^2V^2_{ud}V^2_{us}}{8\pi^5}\sum_{e_1e_2e_3e_4} 
\int_{p_1p_2p_3p_4} \,I(p_1,p_2,p_3,p_4)
\,\tilde{B}^{e_1}_1\tilde{B}^{e_2}_2\tilde{B}^{e_3}_3\tilde{B}^{e_4}_4 \, 
\delta(e_1\tilde{\ve}_1+e_2\tilde{\ve}_2-e_3\tilde{\ve}_3-e_4\tilde{\ve}_4-\d\m) \non
&&\times\,\left[f(-e_1\tilde{\ve}_1)f(-e_2\tilde{\ve}_2)f(e_3\tilde{\ve}_3)f(e_4\tilde{\ve}_4)
-f(e_1\tilde{\ve}_1)f(e_2\tilde{\ve}_2)f(-e_3\tilde{\ve}_3)f(-e_4\tilde{\ve}_4)\right] \, ,
\eea
and for the contribution of the anomalous propagators
\bea \label{tilde2}
\widetilde{\G}_d^{r_1r_2r_3r_4}&\equiv& \frac{G_F^2V^2_{ud}V^2_{us}}{16\pi^5} \sum_{e_1e_2e_3e_4} 
\int_{p_1p_2p_3p_4} \,\tilde{I}(p_1,p_2,p_3,p_4)
\,\frac{e_1\Delta}{2\tilde{\ve}_1}
\tilde{B}_2^{e_2}\tilde{B}_3^{e_3}\frac{e_4\Delta}{2\tilde{\ve}_4} 
\,\delta(e_1\tilde{\ve}_1+e_2\tilde{\ve}_2-e_3\tilde{\ve}_3-e_4\tilde{\ve}_4-\d\m) \non
&&\times\,\left[f(-e_1\tilde{\ve}_1)f(-e_2\tilde{\ve}_2)f(e_3\tilde{\ve}_3)f(e_4\tilde{\ve}_4)
-f(e_1\tilde{\ve}_1)f(e_2\tilde{\ve}_2)f(-e_3\tilde{\ve}_3)f(-e_4\tilde{\ve}_4)\right] \, .
\eea
Because of $\tilde{B}^-_i=0$ for $\Delta=0$,
only one of the 16 terms, namely $e_1=\ldots = e_4=1$, survives in the unpaired phase limit and we 
immediately obtain for this case
\bea \label{normalphase}
\G_d^{\rm unp} &=& \frac{9G_F^2V^2_{ud}V^2_{us}}{8\pi^5}\int_{p_1p_2p_3p_4} \,I(p_1,p_2,p_3,p_4)
\,\delta(p_1+p_2-p_3-p_4)\non
&&\hspace{-2cm}\times\,\left\{[1-f(p_1-\mu)][1-f(p_2-\mu)]f(p_3-\mu_s)f(p_4-\mu)
-f(p_1-\mu)f(p_2-\mu)[1-f(p_3-\mu_s)][1-f(p_4-\mu)]\right\} \, ,
\eea
in accordance with Ref.~\cite{Madsen:1993xx}. Note that in the representation of Eq.\ (\ref{afterang1}), 
all 16 subprocesses contribute even in the unpaired phase (of course yielding the same result 
(\ref{normalphase})). It is easy to understand that the quantities in 
Eq.\ (\ref{newbogol}) are more natural to describe the unpaired phase because $\tilde{\ve}_i=p_i-\mu_i$ 
describes particle excitations for all momenta $p_i$. In contrast, $\ve_i=|p_i-\mu_i|$ describes
hole excitations for $p_i<\mu$ and particle excitations for $p_i>\mu_i$. This distinction between 
particles and holes is not possible in the quasiparticle description of the superconducting 
phase. However, also in the 2SC
phase the new representation is instructive since it reveals the processes that 
are only possible in a superconductor/superfluid, namely the 15 subprocesses that contain at least one 
$\tilde{B}^-_i$. They correspond to processes in which three particles coalesce to one particle
(or one particle decays into three) and in which four particles are annihilated (or created). 
This is possible because particle number is not conserved. Particles can be created from or 
deposited into the condensate. Only when there is no condensate, the ``conventional'' process
of two incoming and two outgoing particles is the only possible one. In the next section, we shall 
study the quantitative contribution of these ``exotic'' processes to the total rate.   

As a result of this subsection, let us rewrite the collision integral (\ref{collision}) 
using the following more illustrative notation,
\be \label{collision1}
\G_d = 4\G_d^{e_1e_2e_4}(\Delta,\Delta,\Delta)  
+ 2 \G_d^{e_1e_4}(\Delta,0,\Delta)  + 2\G_d^{e_2}(0,\Delta,0) + \G_d(0,0,0)+ 
4 \widetilde{\G}_d^{e_1e_2e_4}(\Delta,\Delta,\Delta)+ 2\widetilde{\G}_d^{e_1e_4}(\Delta,0,\Delta)  
 \, , 
\ee
where summation over the respective $e_i=\pm$ is implied. We shall also use the shorthand notation
$\G_d(\Delta,\Delta,\Delta)\equiv \G_d^{e_1e_2e_4}(\Delta,\Delta,\Delta)$ etc.\ where no confusion
is possible.
This notation shows explicitly how many gapped modes participate in the respective 
subprocesses. The definitions of the new quantities is clear from comparison with Eq.\ (\ref{collision}),
the sum over $e_1$, $e_2$, $e_4$ originating from Eqs.\ (\ref{tilde1}) and (\ref{tilde2}).
Since the $s$ quark (corresponding to $e_3$) is ungapped, only terms with $e_3=+1$ survive, and  
the superscripts $e_1$, $e_2$, $e_4$ as well as the three arguments of the functions $\G_d$ correspond 
to the quark flavors $d$, $u_1$, and $u_2$. It is clear from the definition that the number of gapped 
modes determines how many of the 16 subprocesses are nonvanishing. This number can be directly read off 
from the notation in Eq.\ (\ref{collision1}), i.e., all $e_i$'s that are shown explicitly have to 
be summed over $\pm1$, while the missing $e_i$ are set to $+1$, since the contribution for $-1$
vanishes.

\section{Evaluation of the rate}
\label{evaluaterate}

The general evaluation of the collision integral has to be done numerically. As a consistency check, let
us derive the analytical result in the unpaired phase for zero temperature. In this case, all Fermi 
distributions in Eq.\ (\ref{normalphase}) become step functions,
\bea \label{normalstep}
\G_d^{{\rm unp},0}&=&\frac{9G_F^2V^2_{ud}V^2_{us}}{8\pi^5}\int_{p_1p_2p_3} \,I(p_1,p_2,p_3,p_1+p_2-p_3)
\,\Theta(p_1+p_2-p_3)\non
&&\hspace{-2cm}\times\,\left[\Theta(p_1-\mu)\Theta(p_2-\mu)\Theta(\mu_s-p_3)\Theta(\mu-p_1-p_2+p_3)
-\Theta(\mu-p_1)\Theta(\mu-p_2)\Theta(p_3-\mu_s)\Theta(p_1+p_2-p_3-\mu)\right] \, .
\eea
Since we assumed that $\d\m=\mu_s-\mu>0$, the second term in square brackets vanishes.
(The first three step functions in this term imply $p_1+p_2-p_3<2\mu-\mu_s$ which, for $\mu_s>\mu$, 
contradicts the fourth one, implying $p_1+p_2-p_3>\mu$.) Physically, this means that 
at zero temperature the process only happens in one direction. This direction is given by the sign 
of $\d\m$.    
The first term in square brackets can be evaluated upon using the definition of the function $I$ in 
Eq.\ (\ref{I}). One finds 
\bea \label{normalzero}
\G_d^{{\rm unp},0}&=&\frac{6}{5\pi^5}G_F^2V^2_{ud}V^2_{us}\int_\mu^{\mu_s} dp_3\int_\mu^{p_3} dp_2
\int_\mu^{p_3-p_2+\mu} dp_1\, (p_1+p_2-p_3)^3\left[(p_1+p_2)^2+3p_3(p_1+p_2)+6p_3^2\right] \non
&=&  \frac{16}{5\pi^5}G_F^2 V_{us}^2V_{ud}^2\left(\mu^5\d\m^3+\frac{5}{16}\mu^4\d\m^4
-\frac{3}{16}\mu^3\d\m^5+\frac{1}{32}\mu^2\d\m^6+\frac{5}{112}\mu\d\m^7-\frac{15}{896}\d\m^8\right) \, ,
\eea
which is in agreement with Ref.~\cite{Madsen:1993xx}. 

\begin{figure*} [ht]
\begin{center}
\hbox{\includegraphics[width=0.45\textwidth]{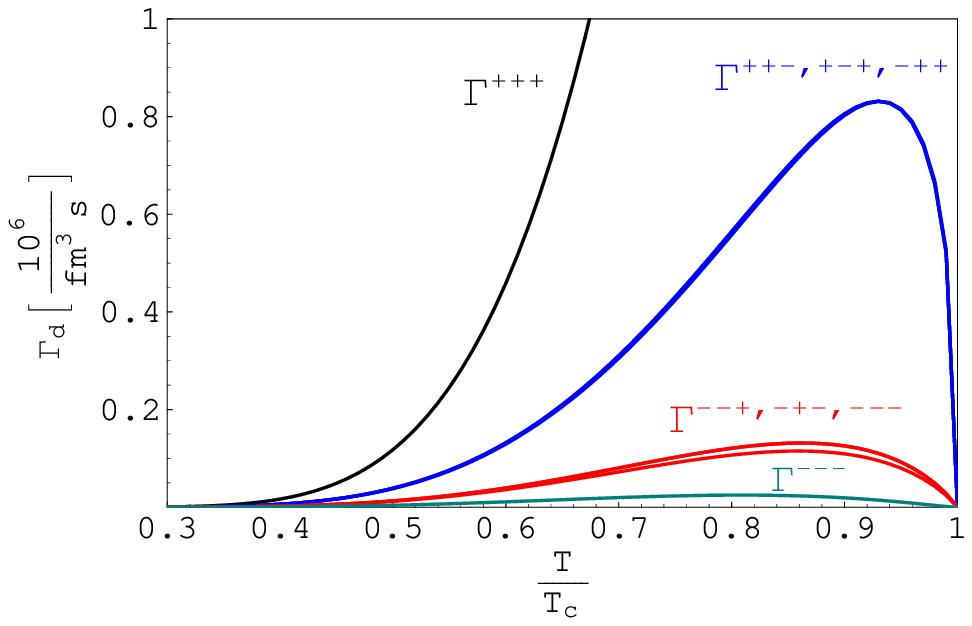}
\hspace{1.5cm}\includegraphics[width=0.45\textwidth]{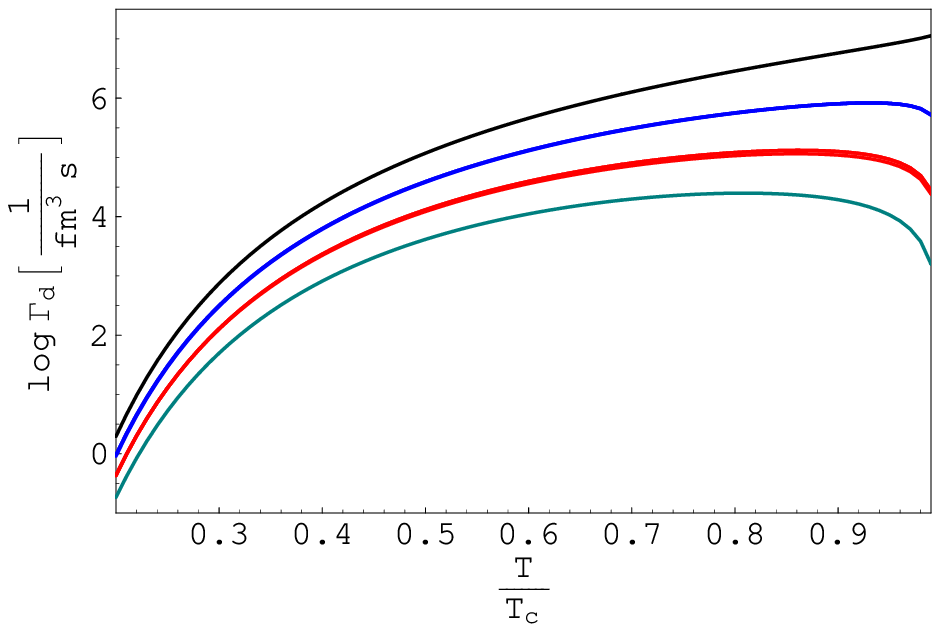}}
%\vspace{0.5cm}
\caption{(Color online) Subprocesses $\Gamma_d^{e_1e_2e_4}(\Delta,\Delta,\Delta)$ on a 
linear scale (left panel) and a logarithmic scale (right panel). The uppermost (black) line corresponds to the process
$\Gamma_d^{+++}$ with two incoming and two outgoing particles. All other processes include creation
or annihilation of particles from the condensate. These processes are only possible in the superconducting 
phase and thus vanish at the critical temperature.} 
\label{fig3gap}
\end{center}
\end{figure*}

We treat all other cases numerically. For the
following results we use $\mu=500$ MeV and $\d\m=10$ MeV. 
We assume a transition temperature of
$T_c=30$ MeV. Via the relation $T_c/\Delta_0=e^\gamma/\pi$, 
where $\gamma\simeq 0.577$ is the Euler-Mascheroni constant, this corresponds
to a zero-temperature gap $\Delta_0\simeq 52.6$ MeV. We shall employ the following model temperature 
dependence of the gap,
\be
\Delta(T) = \Delta_0\sqrt{1-\left(\frac{T}{T_c}\right)^2} \, .
\ee
%From Eq.\ (\ref{collision}) and the discussion in Sec.~\ref{subprocesses}, in particular 
%Fig.~\ref{figdiagrams}, the exact definition and physical meaning of the various terms in 
%Eq.\ (\ref{collision1}) is obvious. 
We have to compute the collision integrals $\G_d(\Delta_1,\Delta_2,\Delta_4)$ 
for all combinations of $\Delta_1$, $\Delta_2$, $\Delta_4$ shown in Eq.\ (\ref{collision1}). 
Let us analyse the various subprocesses before we add them up to the total rate. 
To this end we pick the first term on the right-hand side of Eq.\ (\ref{collision1}), which actually
is a sum of 8 subrates.  We show the rates of all
8 subprocesses of $\G_d(\Delta,\Delta,\Delta)$ in Fig.~\ref{fig3gap}. The 
figure shows that, due to symmetries of the collision integral, some of the resulting curves coincide.
One obtains four groups, characterized by zero, one, two, and three factors $\tilde{B}_i^-$ in the 
collision integral (from top to bottom in the figure). In other words, the three lower curves
in the right panel describe processes of coalescence or decay of particles, while the upper curve
corresponds to a ``normal'' interaction process of two incoming and two outgoing particles.
The left panel shows that this ``normal'' process is the only one that does not 
vanish at the transition temperature. Of course, all other processes can only contribute to the rate 
in the presence of a condensate. Although smaller in magnitude, these processes cannot be completely 
neglected for any temperature range since their contribution increases quickly for temperatures decreasing
down from $T_c$. A quantitative analysis shows that the seven processes where at least one of 
$e_1$, $e_2$, $e_4$  equals $-1$ add up to about 30\% of the total rate at $T/T_c\simeq 0.9$, and their relative contribution increases monotonically with decreasing temperature. 

An evaluation of the 8 subprocesses from Eq.\ (\ref{tilde2}), 
$\widetilde{\G}_d(\Delta,\Delta,\Delta)$,  
shows that their contribution is of the order of the processes $\G_d^{---}(\Delta,\Delta,\Delta)$.
Or, to give another quantitative estimate of these processes, we find that 
$\sum_{e_1e_2e_4}\widetilde{\G}_d^{e_1e_2e_4}(\Delta,\Delta,\Delta) < \G_d^{--+}(\Delta,\Delta,\Delta)$.
Therefore, one can neglect these contributions, i.e., 
we may use $\G_d(\Delta_1,\Delta_2,\Delta_4)$ for the almost identical rate 
$\G_d(\Delta_1,\Delta_2,\Delta_4)+\widetilde{\G}_d(\Delta_1,\Delta_2,\Delta_4)$. 

For the rates $\G_d^{e_1,e_4}(\Delta,0,\Delta)$ and $\G_d^{e_2}(0,\Delta,0)$ there are 4 and 
2 subprocesses, respectively. We do not show the results for these subprocesses explicitly, since  
the corresponding plots are very similar to the ones in Fig.~\ref{fig3gap}.

\begin{figure*} [ht]
\begin{center}
\includegraphics[width=0.45\textwidth]{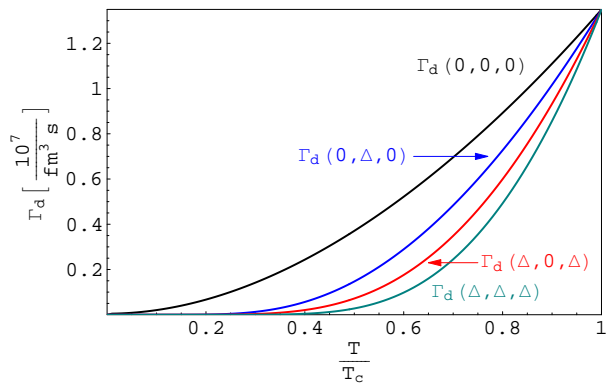}
%\vspace{0.5cm}
\caption{(Color online) Contributions to the 2SC rate, separated by the number of participating 
gapped fermions. The more fermions are gapped, the lower is the rate of the corresponding subprocess.
The lowest (green) curve is the sum of the subprocesses shown in Fig.~\ref{fig3gap}. The sum of all lines
in this figure is the total rate $\G_d$ in the 2SC phase. } 
\label{figseparate2sc}
\end{center}
\end{figure*}

\begin{figure*} [ht]
\begin{center}
\includegraphics[width=0.45\textwidth]{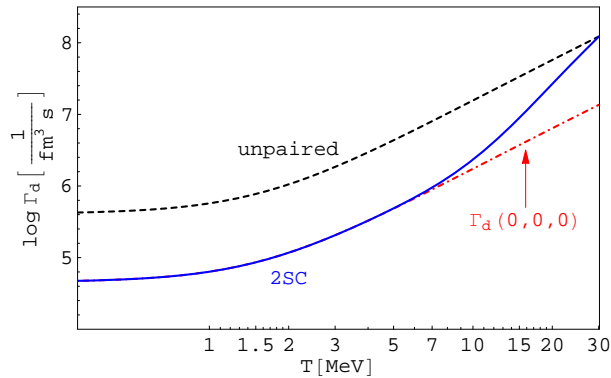}
\caption{(Color online) Total 2SC rate (solid, blue) obtained from Eq.\ (\ref{collision1}) compared to 
the unpaired phase
rate $\G_d^{\rm unp}=9\G_d(0,0,0)$ (dashed, black) and the rate $\G_d(0,0,0)$ (dashed-dotted, red) on a
logarithmic temperature scale (remember $T_c=30$ MeV). For small temperatures, the result approaches 
the zero-temperature result
(which, in the case of the unpaired phase, is given by Eq.\ (\ref{normalzero})). The 2SC rate
is well approximated by $\G_d(0,0,0)$ up to $T\simeq 0.3\, T_c$. Above this temperature, 
effects of the gapped modes are not negligible. For all temperatures,
the rate of the 2SC phase is smaller than that of the unpaired phase. In Sec.\ \ref{calcbulk} we 
shall see that this is not true for the bulk viscosity.} 
\label{fig2sc}
\end{center}
\end{figure*}

After having discussed the subprocesses for a given number of gapped fermions, we shall now compare the
rates for processes with different numbers of gapped fermions. I.e., so far we have picked the first 
of the terms on the right-hand side of Eq.\ (\ref{collision1}) and analyzed the subprocesses hidden in this 
term.
Now we shall compare the different terms in this equation with each other. This is done in  
Fig.~\ref{figseparate2sc}. Every curve in this figure corresponds to the rate of a given number of 
participating gapped fermions, where the number of gapped fermions increases from the top to the bottom 
curve. One sees that 
all rates containing at least one gapped fermion are strongly suppressed for small temperatures.
This suppression can easily be extracted analytically, see Appendix \ref{appsmallT}. 
For $\d\m<\Delta_0$, the leading behavior for small temperatures is 
\be \label{suppression}
\G_d(\Delta,\Delta,\Delta)\sim e^{-(2\Delta-\d\m)/T} \, , \qquad 
\G_d(\Delta,0,\Delta)\sim e^{-\Delta/T} \, , \qquad 
\G_d(0,\Delta,0)\sim e^{-(\Delta-\d\m)/T} \, . \qquad 
\ee
In Appendix \ref{appsmallT}, also the behavior for $\d\m>\Delta_0$, is derived (although this 
regime is not relevant in the present context).
From the suppression due to the gap (\ref{suppression}) we conclude that for small temperatures,
we may approximate the total 2SC rate by the contribution of ungapped fermions, 
$\G_d\simeq \G_d(0,0,0)$ \cite{Madsen:1999ci}. In Fig.~\ref{fig2sc} we see that this 
approximation starts to break down for temperatures $T\gtrsim 10\,{\rm MeV} \simeq 0.3\, T_c$.  

\section{Results for the bulk viscosity}
\label{calcbulk}

In this section, we compute the bulk viscosity of 2SC quark matter as a function of temperature. 
We first discuss the role of the subprocesses in Sec.\ \ref{subcontribution} 
before we compare the total bulk viscosity of the 2SC phase with that of unpaired
quark matter in Sec.\ \ref{finalresults}.
We make use of the expression for the bulk viscosity derived in Sec.~\ref{definebulk},
see Eq.\ (\ref{bulkfinal}). This equation shows that we need the quantities $\omega$, $B$, $C$, and 
$\gamma=C\lambda$. 
For the coefficients $B$ and $C$, see definitions (\ref{defB}) and (\ref{defC}) and Table \ref{tableBC};
we use a strange
mass $m_s=100$ MeV and a chemical potential of $\mu=500$ MeV. In order
to determine $\gamma$, we need the results for the rate $\G_d$, presented in the previous section.
Remember that, in the derivation of the bulk viscosity, leading to Eq.\ (\ref{bulkfinal}), we have assumed
the rate to be linear in $\d\m$, see Eq.\ (\ref{deflambda}). This linear approximation is, for any given 
temperature $T$, valid for a sufficiently small $\d\m$. Physically, this means that we assume the 
volume oscillation $\d V_0/V_0$ to be sufficiently small, because the magnitude of $\d\m$ is proportional to 
$\d V_0/V_0$, see Eq.\ (\ref{amplitude}). Technically, we have to compute the rate $\G_d$ at a sufficiently
small $\d\m$ such that we can determine $\lambda$, and thus the characteristic frequency $\gamma$, 
via $\lambda = \Gamma_d/\d\m$. (Hence, we cannot necessarily use the numerical values from the previous
section which have been obtained for a fixed $\d\m=10$ MeV.) 

\subsection{Contribution of subprocesses to the bulk viscosity}
\label{subcontribution}

\begin{figure*} [ht]
\begin{center}
\includegraphics[width=0.45\textwidth]{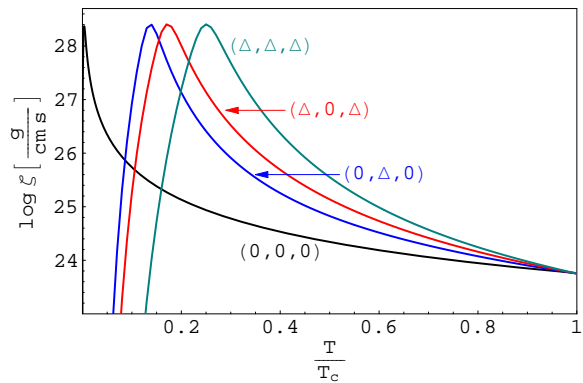}
%\vspace{0.5cm}
\caption{(Color online) Bulk viscosities calculated as if the only 
equilibration processes were those that contribute to the
$\G_d(0,0,0)$ subrate
(black), the $\G_d(0,\Delta,0)$ subrate (blue), 
the $\G_d(\Delta,0,\Delta)$ subrate (red), and
$\G_d(\Delta,\Delta,\Delta)$ subrate (green). We fixed 
$\omega/(2\pi)=10^3{\rm s}^{-1}$ and $\mu=500$ MeV. 
The temperature scale is linear.
}
\label{figbulk0123}
\end{center}
\end{figure*}

Before we turn to the final results for the 2SC phase, let us discuss the contribution of the subprocesses. 
As we have discussed in Sec.\ \ref{evaluaterate}, 
the rates of the subprocesses depend on the number of gapped fermions that participate in the 
process. In Fig.\ \ref{figseparate2sc} we have shown these rates, namely 
$\G_d(\Delta,\Delta,\Delta)$, $\G_d(\Delta,0,\Delta)$, $\G_d(0,\Delta,0)$, and $\G_d(0,0,0)$. 
In Fig.~\ref{figbulk0123}, we show four different bulk viscosities, assuming (fictitiously) that the 
bulk viscosity solely depends on each of these rates. The different suppression of the rates as a function 
of temperature leads to different temperatures at which the bulk viscosity becomes 
maximal (see Sec.\ \ref{finalresults} for a further discussion of the maxima of the viscosity). 
What do we conclude for the 2SC bulk viscosity from these curves? Or, in other words,
how do these curves ``add up'' to yield the total bulk viscosity as a function of temperature? To answer
this question, remember the
analogy between periodically expanding and contracting quark matter and 
an electric circuit with alternating voltage, see Sec.~\ref{definebulk}. In that section 
we explained that 
one may think of a ``capacitance'' that behaves like the inverse of the rate $1/\lambda$, see 
Eq.\ (\ref{replace}).
If the rate consists of a sum of several distinct subrates (as is the case in 2SC quark matter),
small subrates can be neglected if there is a large, dominating subrate. The resulting 
``capacitance'' is then determined by the dominating subrate, just as in an electric 
circuit with capacitors $C_1$, $C_2$ in series where $1/C_{\rm total}=1/C_1+1/C_2$. Therefore, 
the bulk viscosity of 2SC quark matter is dominated by the subprocess of ungapped fermions, $\G_d(0,0,0)$,
although each of the subprocesses with gapped fermions would, for certain temperatures and 
in the absence of $\G_d(0,0,0)$, yield a much {\em larger} bulk viscosity 
(as can be seen by comparing Fig.~\ref{figbulk0123} with the final results, Fig.\ \ref{figbulk}).
Note that it is crucial for this argument that all subprocesses (all ``capacitors'') affect one
single chemical nonequilibrium (one single ``current''), characterized by $\d\m=\mu_s-\mu_d$. 

In Appendix \ref{appstrong} we show how this argument is altered when there is more than one
independent $\d\m$. To illustrate this point let us assume for a moment that the strong interactions, which 
can change the color of the quarks, are no faster than the weak interactions. 
In this case, we must not consider the ``accumulated'' process 
$u+d\leftrightarrow u+s$, but nine processes
$iu+jd\leftrightarrow ju+is$, where $i$ and $j$ each stand for one of three colors, $i,j=red, green, blue$.
And, importantly, there is not only a single $\d\m$, but nine $\d\m_n$'s, one for each of these 
subprocesses. In this picture,
the bulk viscosity is sensitive to the re-equilibration of each of the nine out-of-equilibrium quantities
$\d\m_n$, leading
to a significantly different result. Roughly speaking, the result would be a curve that follows 
all the maxima of the curves in Fig.~\ref{figbulk0123}. However,
this picture is unphysical since it ignores the effect of the strong interactions. They provide 
color-changing 
processes on a time scale much faster than the weak processes. We show in Appendix \ref{appstrong} that,
after taking into account the strong processes, one arrives at the simpler picture
that has been used in Sec.\ \ref{definebulk}: There is only a single out-of-equilibrium
quantity $\d\m$ and the subrates add up like capacitors in series in an electric circuit.
Then, the physical result for the total bulk viscosity is given by Fig.~\ref{figbulk}, 
which we discuss in the following.

\subsection{Bulk viscosity of 2SC quark matter}
\label{finalresults}

\begin{figure*} [ht]
\begin{center}
\hbox{\includegraphics[width=0.45\textwidth]{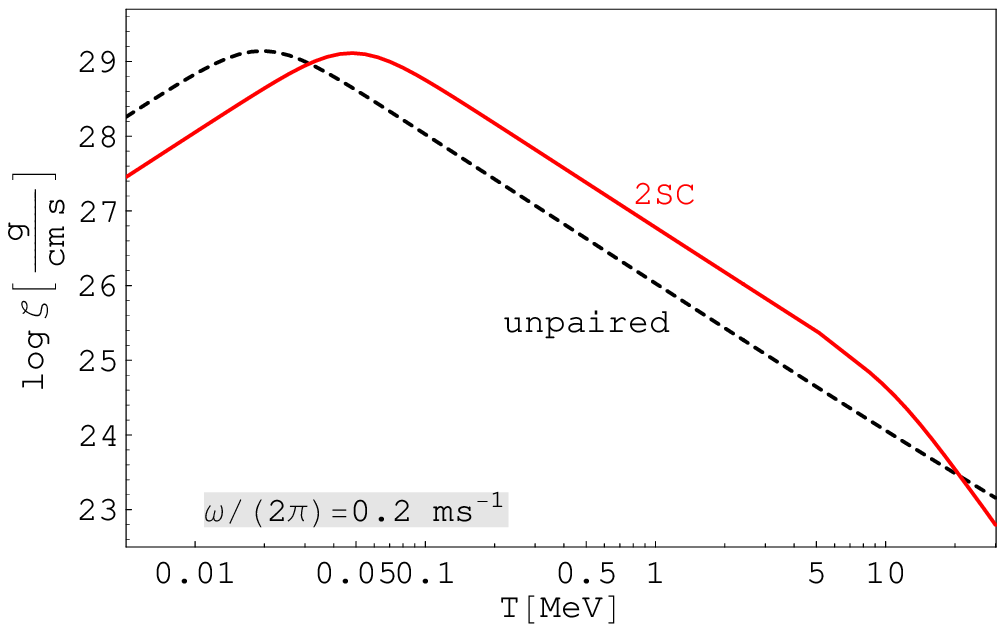}
\hspace{1.5cm}\includegraphics[width=0.45\textwidth]{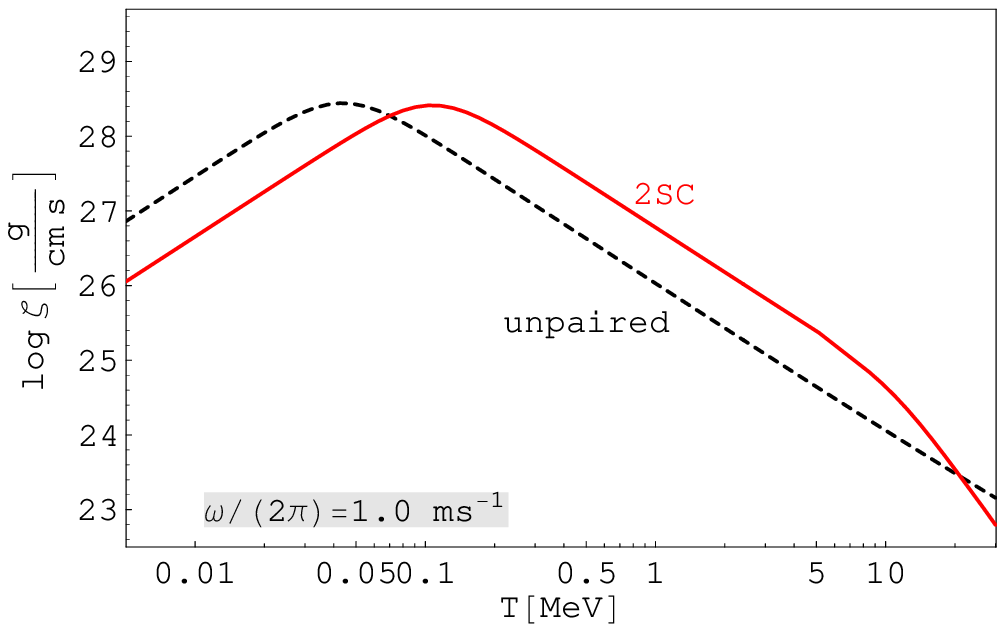}}
\caption{
Bulk viscosities as a function of
temperature for 2SC phase (solid line) and unpaired phase (dashed
line), for a chemical potential $\mu=500$ MeV, and oscillation
frequencies $\omega/(2\pi)=0.2 \,{\rm ms}^{-1}$ (left panel) and
$\omega/(2\pi)=1.0 \,{\rm ms}^{-1}$ (right panel). Note that temperature
is plotted on a logarithmic scale. For $30\,{\rm keV}\lesssim T \lesssim 20\,{\rm MeV}$
and $70\,{\rm keV}\lesssim T \lesssim 20\,{\rm MeV}$, respectively, 2SC has a higher bulk viscosity.}
\label{figbulk}
\end{center}
\end{figure*}

In Fig.\ \ref{figbulk} we show the bulk viscosity for unpaired and 2SC quark matter
for two different oscillation frequencies $\omega/(2\pi)=200\, {\rm
s}^{-1}$ (left panel) and $\omega/(2\pi)=1000\,{\rm s}^{-1}$ (right
panel). We now explain the features of these plots.

%\begin{itemize}{\setlength{\itemsep}{-1ex}}
\begin{list}{$\bullet$}{
  \setlength{\leftmargin}{3ex}
  \setlength{\itemsep}{-0.7\parsep}
  \setlength{\topsep}{-0.7\parskip}
 }

\item The bulk viscosity peaks at a particular temperature. This is
expected from Eq.~(\ref{bulkfinal}): the bulk viscosity as a function
of the rate $\gamma$, for a fixed external frequency $\omega$, has a
maximum at $\gamma=\omega$, when the frequency of the externally
imposed compression matches the microscopic equilibration rate.
Since the microscopic rate $\gamma(T)$ is a
monotonic function of the temperature (see results of the previous
section), we see this maximum when we compute the bulk
viscosity as a function of $T$.

\item The peak viscosities are easily determined by setting $\g=\omega$ in Eq.\ (\ref{bulkfinal}),
\be \label{bulkmax}
\zeta_{\rm max} = \frac{B^2}{2C\,\omega} \, ,
\ee
i.e., they are independent of the microscopic equilibration rate. 
In particular, this simple relation explains the larger maximum values of the viscosity in the 
left panel of Fig.\ \ref{figbulk} compared to the right panel because of the different values of $\omega$.
In general, $\zeta_{\rm max}$ is different for unpaired and 2SC quark matter because the 
values of $B$ and $C$ differ from phase to phase, see Table \ref{tableBC}. For the chosen values
of $m_s$ and $\Delta$, the peak values happen to be almost identical.

\item The 2SC phase has its maximum viscosity at a higher temperature
than the unpaired phase; near and above this temperature (up to $T\simeq 2T_c/3$), the 2SC
phase has a larger viscosity than the unpaired phase.  This follows
from the fact that $\gamma(T)$ is monotonically increasing, but the 2SC
phase has a slower equilibration rate. 
At very low temperatures both rates are
slow, and the higher rate is closer to $\omega$, so the
unpaired phase has higher viscosity \cite{Madsen:1999ci}.  At higher
temperatures, both rates are above $\omega$, but now the slower rate
is closer to $\omega$, so the 2SC phase has higher viscosity. 

\item At low temperature, up to temperatures where the viscosity of the unpaired 
phase peaks, the unpaired and 2SC viscosities differ
by a constant factor (the two lines are parallel on our logarithmic plot).
This follows from $\zeta\propto \gamma$ for $\gamma\ll \omega$ and the results of the
previous section where we have shown that for sufficiently small
temperatures, roughly $T\lesssim 0.3\, T_c$ (see Fig.~\ref{fig2sc}),
the rates simply differ by a factor 9 because the contribution of the
gapped modes can be neglected. Also for temperatures larger than the temperature
at which the viscosity of the 2SC phase peaks up to $T\simeq 10$ MeV, the viscosities differ
by a constant factor. However, they are now in reversed order since $\zeta\propto 1/\gamma$ 
for $\gamma\gg\omega$. 

\item At the critical temperature, here chosen to be $T_c=30$ MeV,
the rates of the two phases become the same (the 2SC gap vanishes). However, due to the ``locking'' 
of Fermi surfaces in the 2SC phase and the resulting difference in the coefficients entering the viscosity,
the viscosities are different. From the values in Table \ref{tableBC} we conclude 
$\zeta_{\rm unp}(T_c)/\zeta_{\rm 2SC}(T_c)=9/4 + {\cal O}(m_s^2/\mu^2)$, in agreement with the results
shown in Fig.\ \ref{figbulk}. 
\end{list}
%\end{itemize}

\section{Conclusions}
\label{conclusions}

We have computed the bulk viscosity of three-flavor color-superconducting quark matter in the 2SC phase 
originating from the nonleptonic weak process $u+d\leftrightarrow u+s$. To this end, we have presented 
a detailed calculation of the rate for this process for arbitrary temperatures. Since 2SC quark matter 
contains ungapped strange quarks as well as gapped and ungapped up and down quarks, 
this process contains several subprocesses. Each of these subprocesses that contains at least one
gapped quasiparticle mode can be further divided into ``subsubprocesses'' which
are similar to the ones in ordinary superfluids such as $^3$He. In both color superconductors
and $^3$He, these subsubprocesses allow for creation and annihilation of quasiparticles from the condensate. Therefore, 
the process $u+d\leftrightarrow u+s$ has contributions from decay and coalescence 
processes. We have presented a formal derivation of the total rate, starting from a kinetic equation 
and making use of the microscopic fermion propagators, resulting in an expression that 
naturally contains all these contributions. Our results are not only interesting for the special case
of the 2SC phase but also, because of the multitude of subprocesses, provide a general quantitative
analysis for the effect of a fermion condensate on the considered nonleptonic process. 

We have evaluated the rate numerically, observing the expected exponential suppressions for small 
temperatures. For subprocesses with three gapped modes we find a suppression of the form 
$e^{-2\Delta/T}$, while two or one gapped mode(s) yield $e^{-\Delta/T}$. 
Upon evaluating the collision integral for temperatures up to the superconducting transition temperature,
we have also studied the temperature regime where the exponential functions are no longer valid approximations
for the rate. 

The results for the rate, which translate into a typical time scale of
the weak process, have been used to compute the bulk viscosity as a
function of temperature for a given external oscillation frequency,
chosen to be of order 1~kHz, which corresponds to the fastest
rotation rate observed for pulsars.
We have pointed out that the bulk viscosity
is the response of the quark system to this external oscillation just
as an electric circuit responds to an external alternating voltage by
an induced current.  In this sense, we have identified a
``resistance'' and a ``capacitance'' of the quark system in terms of
the microscopic process. Using this analogy, it is easy to understand
that the behavior of the bulk viscosity assumes a maximum when the 
microscopic time scale matches that of the external frequency. 

%Our results hold under the assumption that the leptonic processes
%$u+e\leftrightarrow d+ \nu$ and $u+e\leftrightarrow s+ \nu$ are negligible
%for the calculation of the bulk viscosity. For sufficiently small temperatures, 
%where the equilibration rates for both nonleptonic and leptonic processes are 
%much larger than the external frequency, this assumption seems reasonable. However,
%for larger temperatures one has to confirm this assumption by a microscopic
%calculation. Moreover, it remains to refine the current calculation by
%taking into account the strange quark mass for the microscopic rate as
%well as neutrality conditions for the system; both effects are
%expected to slightly modify the bulk viscosity
%quantitatively. 

Our calculations show that, for a transition temperature of $T_c=30$ MeV
and oscillation frequencies in the kHz range,
the bulk viscosity of 2SC quark matter is larger than that of unpaired
quark matter over a wide temperature range. This range is approximately
given by $10^{-3}\,T_c < T < 2T_c/3$ for an oscillation period of one millisecond.
For longer oscillation periods, this range is even larger.
As discussed above, this is because in that temperature
range the slower equilibration rate
of 2SC matter resonates better with kHz frequency oscillations
than does the faster equilibration rate of unpaired quark matter.
This result has potential physical implications for the behavior
of the star early in its life, before its internal temperature drops 
below about 0.1 MeV.
It is also interesting to note that at these early times
neutrinos are trapped, and the resultant lepton number chemical
potential can favor a 2SC phase with strange quarks \cite{Steiner:2002gx} (but see also \cite{Ruster:2005ib}).
Moreover, bulk viscosity is likely to be the dominant factor for damping
of potentially unstable $r$-modes in this very temperature regime \cite{Kokkotas:2001ze,Weber:2004kj}.
Although the star cools down below 0.1 MeV within days, 
this short time might potentially be  
long enough for instabilities to grow and spin down the star drastically \cite{Lindblom:2000az}. 
For hot
newly born stars, we have thus found that quark Cooper pairing
actually works in favor of damping of unstable modes.
%Hence, there are important phenomenological consequences of a large
%bulk viscosity at these temperatures.
For much smaller
temperatures, $T\lesssim 0.1$ keV, shear viscosity effects become
important, and Cooper pairing is expected to reduce the shear
viscosity and thus the damping of the unstable modes
\cite{Madsen:1999ci}.  We have found a similar effect for
the bulk viscosity: for sufficiently small temperatures, $T<10^{-3}T_c$,
the bulk viscosity of 2SC quark matter is smaller 
than that of unpaired quark matter.  Besides the damping of unstable
rotational modes, bulk viscosity is also an important quantity for
time scales of radial pulsation of compact stars \cite{Sahu:2001iv}.
Our results can serve as an ingredient for
calculations addressing the viscous damping of these pulsations.

Finally, we emphasize that the results for the bulk viscosity in the 2SC phase cannot be 
extrapolated to the CFL phase. In this phase, all fermions acquire a gap in their energy spectrum and
thus the rate of the process $u+d\leftrightarrow u+s$ will be smaller than in the 2SC phase. In particular,
this rate will be exponentially suppressed for small temperatures. However, 
kaons (which have nonzero strangeness) serve as a more effective source for re-equilibration, and
their contribution to the bulk viscosity has to be taken into account \cite{preparation}.

\begin{acknowledgments}
The authors like to thank M.\ Braby, P.\ Jaikumar, S.\ Reddy, D.\ Rischke, B.\ Sa'd, 
and I.\ Shovkovy for valuable discussions
and acknowledge support by the U.S. Department of Energy
under contracts 
\#DE-FG02-91ER40628  % Wash U theory
and
\#DE-FG02-05ER41375 (OJI). % Mark's OJI
\end{acknowledgments}

\appendix

\section{Bulk viscosity including leptonic processes}
\label{appleptonic}

In this appendix, we derive an expression for the bulk viscosity which is valid for a system that
not only converts $s$ into $u$ quarks via the non-leptonic process considered in the main part of the 
paper, but also contains leptonic processes. We shall consider the following three processes, employing
the linear approximation for the several rates as in Sec.~\ref{definebulk},
\begin{subequations} \label{123}
\bea
u+d&\leftrightarrow &u+s \, , \qquad \G_1=\lambda_1 \d\m_1 \, , \\
u+e &\leftrightarrow &d+\nu \, , \qquad \G_2=\lambda_2 \d\m_2 \, , \\
u+e &\leftrightarrow &s+\nu \, , \qquad \G_3=\lambda_3 \d\m_3 \, .
\eea
\end{subequations}
The quantities $\G_1$, $\lambda_1$, and $\d\m_1$ correspond to $\G_d$, $\lambda$, and $\d\m$ in Sec.\ 
\ref{definebulk}. In chemical equilibrium, $\d\m_1=\d\m_2=\d\m_3=0$. Out of equilibrium, the quantities
\begin{subequations}
\bea
\d\m_1&=&\mu_s-\mu_d \, , \\
\d\m_2&=&\mu_d-\mu_u-\mu_e \, , \\
\d\m_3&=&\mu_s-\mu_u-\mu_e=\d\m_1+\d\m_2 \, , 
\eea
\end{subequations}
are nonzero (we assume $\mu_\nu=0$). Note that there are only two independent $\d\m$'s. According to the 
above microscopic processes, the densities for electrons and quarks change as follows,
\begin{subequations}\label{dndt}
\bea
\frac{dn_e}{dt}&=&\G_2+\G_3=\lambda_3\d\m_1+(\lambda_2+\lambda_3)\d\m_2 \, , \\
\frac{dn_d}{dt}&=&\G_1-\G_2=\lambda_1\d\m_1-\lambda_2\d\m_2 \, , \\
\frac{dn_s}{dt}&=&-\G_1-\G_3=-\frac{dn_d}{dt}-\frac{dn_e}{dt} \, , \\
\frac{dn_u}{dt}&=&\G_2+\G_3=\frac{dn_e}{dt} \, . 
\eea
\end{subequations}
We can thus express all density changes in terms of the changes in $n_e$ and $n_d$.
Analogous to Sec.~\ref{definebulk}, we write the pressure as
\bea
p(t)&=& p_0+\left(\frac{\partial p}{\partial V}\right)_0\d V +
\left(\frac{\partial p}{\partial n_e}\right)_0\d n_e+ 
\left(\frac{\partial p}{\partial n_d}\right)_0\d n_d+\left(\frac{\partial p}{\partial n_s}\right)_0\d n_s+  
\left(\frac{\partial p}{\partial n_u}\right)_0\d n_u \non
&=&p_0+\left(\frac{\partial p}{\partial V}\right)_0\d V + (B_1+B_2)\d n_e + B_1 \d n_d \, , 
\eea
where 
\begin{subequations}
\bea
B_1&\equiv& \left(n_d\frac{\partial \mu_d}{\partial n_d}\right)_0- 
\left(n_s\frac{\partial \mu_s}{\partial n_s}\right)_0 \, , \\
B_2&\equiv& \left(n_u\frac{\partial \mu_u}{\partial n_u}\right)_0+ 
\left(n_e\frac{\partial \mu_e}{\partial n_e}\right)_0- 
\left(n_d\frac{\partial \mu_d}{\partial n_d}\right)_0 \, , 
\eea
\end{subequations}
and we have made use of the relations between the various densities given in Eqs.\ (\ref{dndt}).
For the purpose of this appendix, it is sufficient to consider unpaired quark matter. In the 2SC
phase, $B_1$ and $B_2$ would have additional terms due to the pairing of $u$ and $d$ quarks. Since the 
conclusions of this appendix do not depend on the values of $B_1$ and $B_2$ we may for simplicity proceed 
without taking into account the effects of pairing.
Then, the average power dissipation, cf.\ Eq.\ (\ref{resultpower}), is
\be \label{powermulti}
\left\langle\frac{dW}{dt}\right\rangle=a_1\,\langle\d\m_1(t)\d v(t)\rangle
+a_2\,\langle\d\m_2(t)\d v(t)\rangle \, ,
\ee
where we abbreviated
\begin{subequations}
\bea
a_1&\equiv& B_1\lambda_1+(B_1+B_2)\lambda_3 \, , \\
a_2&\equiv& B_2\lambda_2+(B_1+B_2)\lambda_3 \, .
\eea
\end{subequations}
The changes in the chemical potentials are derived as in Sec.~\ref{definebulk}, cf.\ Eqs.\ (\ref{deltamu})
and (\ref{diffeq}), 
\be \label{twodiffeqs}
\frac{\partial \d\mu_i}{\partial t}= B_i \frac{\partial \d v}{\partial t} - \alpha_i
\,\d\m_1 - \beta_i\,\d\m_2 \, , \qquad i=1,2 \, ,
\ee
where
\begin{subequations}
\bea
\alpha_i&\equiv& C_i^d\lambda_1+C_i^e\lambda_3 \, , \\
\beta_i&\equiv& C_i^e(\lambda_2+\lambda_3)-C_i^d\lambda_2 \, , 
\eea
\end{subequations}
with
\begin{subequations}
\bea
C_1^d&\equiv& \left(\frac{\partial\mu_d}{\partial n_d}\right)_0
+\left(\frac{\partial\mu_s}{\partial n_s}\right)_0 \, , \qquad
C_1^e\equiv \left(\frac{\partial\mu_s}{\partial n_s}\right)_0 \, , \\
C_2^d&\equiv& -\left(\frac{\partial\mu_d}{\partial n_d}\right)_0\, , \qquad 
C_2^e\equiv \left(\frac{\partial\mu_u}{\partial n_u}\right)_0
+\left(\frac{\partial\mu_e}{\partial n_e}\right)_0 \, . 
\eea
\end{subequations}
The single differential equation (\ref{diffeq}) thus has become a system of two coupled 
differential equations (\ref{twodiffeqs}). With the ansatz $\d\mu_i={\rm Re}(\d\m_i^0 e^{i\omega t})$
we may reduce the system of differential equations to an algebraic set of four linear equations
for the four variables ${\rm Re}\, \d\m_1^0$, ${\rm Im}\, \d\m_1^0$, ${\rm Re}\, \d\m_2^0$, 
${\rm Im}\, \d\m_2^0$,
\begin{subequations}
\bea
\omega\, {\rm Re}\, \d\m_i^0&=&\omega B_i\,\d v_0 - \alpha_i\,{\rm Im}\,\d\m_1^0
- \beta_i\, {\rm Im}\,\d\m_2^0 \, , \\
\omega\, {\rm Im}\,\d\m_i^0&=& \alpha_i\,{\rm Re}\,\d\m_1^0
+ \beta_i\, {\rm Re}\,\d\m_2^0 \, .
\eea
\end{subequations}
From Eq.\ (\ref{powermulti}) and $\d v(t)=\d v_0\cos\omega t$ we conclude that only the real parts
of the complex amplitudes $\d\m_i^0$ enter the power dissipation. They are given by
\begin{subequations}
\bea
{\rm Re}\,\d\m_1^0&=&\d v_0\,\omega^2\frac{B_1(\b_2^2+\a_2\b_1)-B_2\b_1(\a_1+\b_2)+B_1\omega^2}
{(\a_2\b_1-\a_1\b_2)^2+(2\a_2\b_1+\a_1^2+\b_2^2)\omega^2+\omega^4} \, , \\
{\rm Re}\,\d\m_2^0&=&\d v_0\,\omega^2\frac{B_2(\a_1^2+\a_2\b_1)-B_1\a_2(\a_1+\b_2)+B_2\omega^2}
{(\a_2\b_1-\a_1\b_2)^2+(2\a_2\b_1+\a_1^2+\b_2^2)\omega^2+\omega^4} \, .
\eea
\end{subequations}
Consequently, Eq.\ (\ref{powermulti}) and the definition (\ref{defbulk}) yield the bulk viscosity
\be \label{multibulk}
\zeta = a_1 \frac{{\rm Re}\, \d\m_1^0}{\d v_0\,\omega^2} + a_2 \frac{{\rm Re}\, \d\m_2^0}{\d v_0\,\omega^2} = 
\frac{P_1 + P_2\,\omega^2}{Q_1 + Q_2\,\omega^2 +\omega^4} \, ,
\ee
where
\begin{subequations} \label{PQ}
\bea 
P_1&\equiv& (\lambda_1\lambda_2+\lambda_1\lambda_3+\lambda_2\lambda_3)\Big(B_1^2\left\{(C_2^d-C_2^e)
[C_2^d\lambda_2-C_2^e(\lambda_2+\lambda_3)]+(C_1^e-C_1^d)(C_2^d\lambda_1+C_2^e\lambda_3)\right\} \non
&& +\, B_1B_2\left\{(C_1^d-C_1^e-C_2^d)[C_1^d\lambda_1+(C_2^e-C_2^d)\lambda_2]+[C_1^e(C_1^d-C_1^e+C_2^d)-C_2^e(C_1^d+C_1^e+C_2^d)]\lambda_3\right\} \non
&&+\, B_2^2[(C_1^d)^2\lambda_1+(C_1^e-C_1^d)C_2^d\lambda_2+(C_1^d+C_2^d)C_1^e\lambda_3]\Big)\, , \\
P_2&\equiv& B_1^2\lambda_1+B_2^2\lambda_2+(B_1+B_2)^2\lambda_3 \, , \\
Q_1&\equiv& (\lambda_1\lambda_2+\lambda_1\lambda_3+\lambda_2\lambda_3)^2(C_1^eC_2^d-C_1^dC_2^e)^2 \, , \\
Q_2&\equiv& (C_1^d\lambda_1+C_1^e\lambda_3)^2 + 2(C_2^d\lambda_1+C_2^e\lambda_3)[C_1^e(\lambda_2+\lambda_3)-C_1^d\lambda_2]+[C_2^d\lambda_2-C_2^e(\lambda_2+\lambda_3)]^2 \, . 
\eea
\end{subequations}
With the densities (of unpaired quark matter) 
$n_{u/d}=\mu_{u/d}^3/\pi^2$, $n_s=(\mu_s^2-m_s^2)^{3/2}/\pi^2$, and $n_e=\mu_e^3/(3\pi^2)$
we have
\begin{subequations}
\bea
B_1&=&\frac{m_s^2}{3\mu_s} \, , \qquad B_2 = 0 \, , \\
C_1^d&=&\frac{\pi^2}{3\mu_s^2}\left(1+\frac{\mu_s}{\sqrt{\mu_s^2-m_s^2}}\right) \,  , \qquad
C_1^e=\frac{\pi^2}{3\mu_s\sqrt{\mu_s^2-m_s^2}} \, , \\
C_2^d&=&-\frac{\pi^2}{3\mu_s^2} \, , \qquad 
C_2^e = \frac{\pi^2}{3}\left(\frac{1}{\mu_u^2}+\frac{3}{\mu_e^2}\right)
\, . 
\eea
\end{subequations}
From the definitions it it clear that $B_1$ and $C_1^d$ correspond to $B$ and $C$ in Sec.~\ref{definebulk},
respectively. The result (\ref{multibulk}) together with the definitions (\ref{PQ}) shows
that the three processes are entangled in a complicated way. Let us consider some limit
cases in which the processes disentangle. In particular, we are interested in the case $\lambda_1\gg
\lambda_2, \lambda_3$, since we expect the rates for the leptonic processes to be much smaller
than the rate for the non-leptonic process. In this case,
\be
\zeta\simeq B_1^2\lambda_1\frac{\lambda_1(\lambda_2+\lambda_3)(C_1^d-C_1^e)C_1^d+\omega^2}{\lambda_1^2
(\lambda_2+\lambda_3)^2(C_1^eC_2^d-C_1^dC_2^e)^2+(C_1^d\lambda_1)^2\omega^2+\omega^4} \, .
\ee
In order to obtain some qualitative insight 
we have to compare the inverse time scale $\omega$ with the inverse time scales set by $\lambda_1$,
$\lambda_2$, and $\lambda_3$. For simplicity, let us set $\lambda_2=\lambda_3$.
As a rough estimate, we shall assume that the factors containing one or several 
$C$'s all contribute by a factor $1/\mu_s^2$, i.e., $C_1^d-C_1^e\sim C_1^d\to 1/\mu_s^2$, 
$C_1^eC_2^d-C_1^dC_2^e\sim (C_1^d)^2 \to 1/\mu_s^4$. Then, the relevant inverse time scales 
\be
\gamma_1\equiv \frac{\lambda_1}{\mu_s^2} \, , \qquad \gamma_2\equiv \frac{\lambda_2}{\mu_s^2}
\ee
are, for fixed $\mu_s$, indeed set solely by the corresponding rates of the non-leptonic and 
leptonic processes, respectively, and we can write 
\be
\zeta \sim \g_1\frac{\g_1\g_2+\omega^2}{\g_1^2\g_2^2+\g_1^2\omega^2+\omega^4} \, .
\ee
Obviously, it is incorrect to neglect the much slower process in general. Which process can be neglected
rather depends on $\omega$. In particular, we may consider several limit cases in which the time scales
clearly separate, see Table \ref{tableg1g2}. The general and plausible conclusion from this table
is that the bulk viscosity is dominated by the process whose time scale is closest to the time scale set
by the external oscillation. (Note that this is in contrast to subprocesses which contribute to 
one single $\delta\mu$, as considered in the main part of the paper. In this case, the fastest
subprocess dominates the bulk viscosity even if there is a slower subprocess which, in the 
absence of the fast one, would produce a much larger bulk viscosity.)
More specifically, the leptonic processes can be neglected in the high frequency limit (first row of
Table \ref{tableg1g2}) or  when the frequency is close to the rate of the non-leptonic process
(second row). When the frequency is smaller than the rate of the nonleptonic process, the leptonic
processes can only be neglected if its rate is much smaller than the frequency {\em and} 
$\omega/\g_1\gg\g_2/\omega$, i.e., if the frequency is much closer to the rate of the non-leptonic
process than to the rate of the leptonic processes.   
\begin{table}  
\begin{tabular}[t]{|c|c|}
\hline
$\;\;$ order of time scales$\;\;$& $\zeta$ \\ \hline\hline
$\omega\gg\g_1\gg\g_2$ & $\g_1/\omega^2$ \\ \hline
$\;\;\omega\simeq\g_1\gg\g_2\;\;$ & $\g_1/(\g_1^2+\omega^2)$ \\ \hline
$\g_1\gg\omega\gg\g_2$ & $\;\;\g_2/\omega^2+1/\g_1\;\;$ \\ \hline
$\g_1\gg\omega\simeq\g_2$ & $\g_2/(\g_2^2+\omega^2)$ \\ \hline
$\g_1\gg\g_2\gg\omega$ & $1/\g_2$ \\ \hline
\end{tabular}
\caption{Qualitative behavior of the bulk viscosity $\zeta$ for different limit cases in which the time scales
given by the external oscillation frequency $\omega$, the non-leptonic process $\g_1$, and the leptonic 
process $\g_2$ are clearly separated.}
\label{tableg1g2}
\end{table}

\section{Bulk viscosity including strong processes}
\label{appstrong}

In this appendix, we show that the expression for the bulk viscosity derived in 
Sec.~\ref{definebulk} is only correct in the presence of strong processes which are infinitely
fast compared to the weak process $u+d\leftrightarrow u +s$. The treatment in Sec.~\ref{definebulk}
apparently simplifies the situation because the process $u+d\leftrightarrow u +s$ in fact 
includes several combinations of quark colors. In principle, 
one has to treat all possible processes in a way analogous to the method presented in 
Appendix \ref{appleptonic}, i.e., we have to consider the processes
\begin{subequations} \label{nineprocesses}
\bea
bu+bd&\leftrightarrow &bu+bs \, , \qquad \G_1=\lambda_1\, \d\m_1 \, , \\
gu+bd&\leftrightarrow &bu+gs \, , \qquad \G_2=\lambda_2\, \d\m_2 \, , \\
ru+bd&\leftrightarrow &bu+rs \, , \qquad \G_3=\lambda_3\, \d\m_3 \, , \\
bu+gd&\leftrightarrow &gu+bs \, , \qquad \G_4=\lambda_4\, \d\m_4 \, , \\
bu+rd&\leftrightarrow &ru+bs \, , \qquad \G_5=\lambda_5\, \d\m_5 \, , \\
gu+gd&\leftrightarrow &gu+gs \, , \qquad \G_6=\lambda_6\, \d\m_6 \, , \\
ru+gd&\leftrightarrow &gu+rs \, , \qquad \G_7=\lambda_7\, \d\m_7 \, , \\
gu+rd&\leftrightarrow &ru+gs \, , \qquad \G_8=\lambda_8\, \d\m_8 \, , \\
ru+rd&\leftrightarrow &ru+rs \, , \qquad \G_9=\lambda_9\, \d\m_9 \, . 
\eea
\end{subequations}
As in Eqs.\ (\ref{123}) we assign a rate $\G_i$ to each of the processes and assume these rates 
to be proportional to the corresponding $\d\mu_i$. There are nine different processes, because
we can assign one of three colors ($r$, $g$, $b$) to each of the two (color-conserving) vertices of the process
$u+d\leftrightarrow u +s$. From the results in Secs.~\ref{deriverate} and \ref{evaluaterate} we know that 
not all of the subprocesses have different rates. Due to the remaining $SU(2)$ color symmetry, we rather 
have four different rates depending on the number of gapped quasiparticles participating in the process.
Using the notation introduced in Eq.\ (\ref{collision1}), we have 
\begin{subequations} 
\bea
\G_1&=&\G_d(0,0,0) \, , \\
\G_2&=&\G_3=\G_d(0,\Delta,0) \, , \\
\G_4&=&\G_5=\G_d(\Delta,0,\Delta) \, , \\
\G_6&=&\G_7=\G_8=\G_9=\G_d(\Delta,\Delta,\Delta) \, . 
\eea
\end{subequations}
However, for the purpose of this appendix we do not have to make any assumptions for the rates and may treat 
all nine rates, given by $\lambda_1, \ldots, \lambda_9$, as independent. 

The various $\d\m_i$'s can be written in terms of nine chemical potentials 
(one for each combination of color and flavor). It turns out that only five of them are independent,
\begin{subequations}\label{nine}
\bea
\d\m_1&=&\mu_{bs}-\mu_{bd} \, , \\
\d\m_2&=&\mu_{bu}+\mu_{gs}-\mu_{gu}-\mu_{bd} \, , \\
\d\m_3&=&\mu_{bu}+\mu_{rs}-\mu_{ru}-\mu_{bd}=\d\mu_2-\d\m_6+\d\m_7 \, , \\
\d\m_4&=&\mu_{gu}+\mu_{bs}-\mu_{bu}-\mu_{gd}=\d\m_1-\d\m_2+\d\m_6 \, , \\
\d\m_5&=&\mu_{ru}+\mu_{bs}-\mu_{bu}-\mu_{rd}=\d\m_1-\d\m_2+\d\m_6-\d\m_7+\d\m_9 \, , \\
\d\m_6&=&\mu_{gs}-\mu_{gd} \, , \\
\d\m_7&=&\mu_{gu}+\mu_{rs}-\mu_{ru}-\mu_{gd} \, , \\
\d\m_8&=&\mu_{ru}+\mu_{gs}-\mu_{gu}-\mu_{rd}=\d\m_6-\d\m_7+\d\m_9 \, , \\
\d\m_9&=&\mu_{rs}-\mu_{rd} \, .
\eea
\end{subequations}
We may thus choose $\d\m_1$, $\d\m_2$, $\d\m_6$, $\d\m_7$, $\d\m_9$, as independent variables.
The density changes due to the above processes can be read off from Eqs.\ (\ref{nineprocesses}),
\begin{subequations}
\bea
\frac{dn_{ru}}{dt}&=&\G_3-\G_5+\G_7-\G_8 \, , \qquad 
\frac{dn_{rd}}{dt}=\G_5+\G_8+\G_9 \, , \qquad
\frac{dn_{rs}}{dt}=-\G_3-\G_7-\G_9 \, , \\
\frac{dn_{gu}}{dt}&=&\G_2-\G_4-\G_7+\G_8 \, , \qquad
\frac{dn_{gd}}{dt}=\G_4+\G_6+\G_7 \, , \qquad
\frac{dn_{gs}}{dt}=-\G_2-\G_6-\G_8 \, , \\
\frac{dn_{bu}}{dt}&=&-\G_2-\G_3+\G_4+\G_5 \, , \qquad
\frac{dn_{bd}}{dt}=\G_1+\G_2+\G_3 \, , \qquad
\frac{dn_{bs}}{dt}=-\G_1-\G_4-\G_5 \, . 
\eea
\end{subequations}
Obviously, the total change of the $u$ flavor density vanishes, while the total change of 
$d$ and $s$ flavor densities has the same absolute value but opposite signs,
\be \label{flavorrates}
\frac{dn_u}{dt} = 0 \, , \qquad \frac{dn_d}{dt}=-\frac{dn_s}{dt}=\sum_{i=1}^9\G_i 
\, ,
\ee
where
\be\label{flavordensities}
n_i \equiv n_{ri} + n_{gi} + n_{bi} \, , \qquad i = u,d,s \, .
\ee
If the bulk viscosity was determined solely by the above nine subprocesses of the weak process 
$u+d\leftrightarrow u+s$, we would have to set up a set of five coupled differential equations
for the variables $\d\m_1$, $\d\m_2$, $\d\m_6$, $\d\m_7$, $\d\m_9$, analogous to Eq.\ (\ref{twodiffeqs}). 
Its solution would 
yield the bulk viscosity through an equation analogous to Eq.\ (\ref{powermulti}). We have checked that
this procedure would yield a result that differs from the one presented in 
Sec.~\ref{calcbulk}. In particular: While the bulk viscosity given by Eq.\ (\ref{bulkfinal})
is dominated by the fastest of the subprocesses (here $\lambda_1,\ldots ,\lambda_9$), the setup presented
above would potentially allow all subprocesses to contribute 
significantly to the bulk viscosity (the dominant subprocesses would be determined by the external 
oscillation frequency $\omega$). In other words, this setup requires all chemical nonequilibria  
to re-equilibrate separately, and the one that resonates best with the external oscillation 
gives the largest contribution. In the remainder of this 
appendix, however, we show that this picture is invalidated by the presence of the strong interactions 
which allows the fermions to change their color on a much faster time scale than the one given by the 
weak interaction. 

The relevant strong processes change the colors of two different quark flavors. Since there are three 
pairs of distinct flavors and three pairs of distinct
colors, we have to consider the nine processes
\begin{subequations}
\bea
ru+gd&\leftrightarrow &gu+rd \, , \qquad ru+gs\leftrightarrow gu+rs \, , \qquad 
rd+gs\leftrightarrow gd+rs \, ,  \\
ru+bd&\leftrightarrow &bu+rd \, , \qquad ru+bs\leftrightarrow bu+rs \, , \qquad 
rd+bs\leftrightarrow bd+rs \, ,  \\
gu+bd&\leftrightarrow &bu+gd \, , \qquad gu+bs\leftrightarrow bu+gs \, , \qquad 
gd+bs\leftrightarrow bd+gs \, . 
\eea
\end{subequations}
Assuming these processes to be infinitely fast (this is equivalent to the assumption that the system 
be in chemical equilibrium with respect to all of these processes at any time) 
translates into the following conditions for the chemical potentials,
\begin{subequations}
\bea
\mu_{ru}+\mu_{gd}-\mu_{gu}-\mu_{rd}&=&0 \, , \qquad \mu_{ru}+\mu_{gs}-\mu_{gu}-\mu_{rs}=0 \, , \qquad 
\mu_{rd}+\mu_{gs}-\mu_{gd}-\mu_{rs}=0 \, ,  \\
\mu_{ru}+\mu_{bd}-\mu_{bu}-\mu_{rd}&=&0 \, , \qquad \mu_{ru}+\mu_{bs}-\mu_{bu}-\mu_{rs}=0 \, , \qquad 
\mu_{rd}+\mu_{bs}-\mu_{bd}-\mu_{rs}=0 \, ,  \\
\mu_{gu}+\mu_{bd}-\mu_{bu}-\mu_{gd}&=&0 \, , \qquad \mu_{gu}+\mu_{bs}-\mu_{bu}-\mu_{gs}=0 \, , \qquad 
\mu_{gd}+\mu_{bs}-\mu_{bd}-\mu_{gs}=0 \, .  
\eea
\end{subequations}
These equations result in four independent constraints. Consequently, we may express four chemical 
potentials in terms of the other five, e.g., 
\begin{subequations} \label{fourfive}
\bea
\mu_{gd}&=&\mu_{rd}+\mu_{gu}-\mu_{ru} \, , \qquad \mu_{gs}=\mu_{rs}+\mu_{gu}-\mu_{ru} \, , \\
\mu_{bd}&=&\mu_{rd}+\mu_{bu}-\mu_{ru} \, , \qquad \mu_{bs}=\mu_{rs}+\mu_{bu}-\mu_{ru} \, . 
\eea
\end{subequations}
Inserting these equations into Eqs.\ (\ref{nine}) reduces the number of independent 
$\d\m$'s to one,
\be
\d\m_1 = \ldots = \d\m_9= \mu_{rs} - \mu_{rd} \equiv \d\m \, .
\ee
In order to set up the differential equation for this $\d\m$, we have to know the dependence of the
densities on the remaining chemical potentials (after eliminating $\mu_{gd}$, $\mu_{bd}$, $\mu_{gs}$, 
$\mu_{bs}$ via Eqs.\ (\ref{fourfive})). Then, in the 2SC phase, the densities of the paired quarks
$n_{ru}$, $n_{rd}$, $n_{gu}$, $n_{gd}$ depend on the average chemical potential $(\mu_{rd}+\mu_{gu})/2$.
The densities of the unpaired quarks are straightforwardly obtained with Eqs.\ (\ref{fourfive}).
Let us, for simplicity, assume that $\mu_{ru}=\mu_{gu}=\mu_{bu}\equiv\mu_u$. Then, we use 
the flavor densities from Eq.\ (\ref{flavordensities}) to obtain
\bea \label{diffstrong}
\frac{\partial \d\mu}{\partial t}&=&
-\sum_{i=u,d,s}\left(n_i\frac{\partial \d\mu}{\partial n_i}\right)_0\,
\frac{\partial \delta v}{\partial t} + \sum_{i=u,d}
\left(\frac{\partial \d\mu}{\partial n_i}\right)_0\frac{dn_i}{dt} \, , 
\eea
with $dn_i/dt$ given in Eq.\ (\ref{flavorrates}).
The same procedure is applied to the expression for the power dissipation, cf.\ Eq.\ (\ref{resultpower}).
Upon using $\l_1 + \ldots + \l_9 = \lambda$ with $\lambda$ defined in Eq.\ (\ref{deflambda}), 
we see that the final result is identical to Eq.\ (\ref{bulkfinal}).
Consequently, we have proven that, due to (infinitely fast) strong processes, one can ignore the subtleties
related to different $\d\m$'s corresponding to subprocesses of $u+d\leftrightarrow u+s$. One
rather may solely consider the rate $\G_d$ which is the sum of the rates of these subprocesses and a single 
$\d\m$ related to this rate. This is the method used in the main part of the paper.

\section{Dirac traces and contractions}
\label{appdirac}

From the definitions (\ref{diracT}) and (\ref{diracU}) we find  
\begin{subequations}
\bea
{\cal T}^{00}(\uk,\up) &=& 2\,(1+\uk\cdot\up) \,\, , \\
{\cal T}^{0i}(\uk,\up) &=& 2\,[\hk^i+\hp^i+i\,(\uk\times\up)^i] \,\, , \\
{\cal T}^{i0}(\uk,\up) &=& 2\,[\hk^i+\hp^i-i\,(\uk\times\up)^i] \,\, , \\
{\cal T}^{ij}(\uk,\up) &=& 2\,[\d^{ij}(1-\uk\cdot\up)+\hk^i\hp^j+\hk^j\hp^i+i\,\e^{ij\ell}\,
(\hk^\ell - \hp^\ell)] \, , 
\eea
\end{subequations}
and 
\begin{subequations}
\bea
{\cal U}^{00}(\uk,\up) &=& - 2\,(1+\uk\cdot\up) \,\, , \\
{\cal U}^{0i}(\uk,\up) &=& 2\,[\hk^i+\hp^i+i\,(\uk\times\up)^i] \,\, , \\
{\cal U}^{i0}(\uk,\up) &=& - 2\,[\hk^i+\hp^i-i\,(\uk\times\up)^i] \,\, , \\
{\cal U}^{ij}(\uk,\up) &=& 2\,[\d^{ij}(1-\uk\cdot\up)+\hk^i\hp^j+\hk^j\hp^i+i\,\e^{ij\ell}\,
(\hk^\ell - \hp^\ell)] \, . 
\eea
\end{subequations}
Consequently, we obtain the contractions
\begin{subequations}
\bea
{\cal T}^{\m\n}(\up_4,\up_1){\cal T}_{\m\n}(\up_3,\up_2) &=& 16(1-\up_4\cdot\up_3)(1-\up_1\cdot\up_2) \, , \\
{\cal U}^{\n\m}(\up_1,\up_4){\cal T}_{\m\n}(\up_3,\up_2) &=& 8\left\{
(1-\up_4\cdot\up_3)(1-\up_1\cdot\up_2)+(1+\up_1\cdot\up_3)(1+\up_4\cdot\up_2)-(1+\up_1\cdot\up_4)(1+\up_3\cdot\up_2)
\right.\non
&&\left.-\;i\left[(\up_3+\up_2)\cdot(\up_1\times\up_4)+
(\up_1+\up_4)\cdot(\up_2\times\up_3)\right]\right\} \, ,\label{UT1}\\
{\cal U}^{\m\n}(\up_4,\up_1){\cal T}_{\m\n}(\up_3,\up_2) &=& 8\left\{
(1-\up_4\cdot\up_3)(1-\up_1\cdot\up_2)+(1+\up_1\cdot\up_3)(1+\up_4\cdot\up_2)-(1+\up_1\cdot\up_4)(1+\up_3\cdot\up_2)
\right.\non
&&\left.+\;i\left[(\up_3+\up_2)\cdot(\up_1\times\up_4)+
(\up_1+\up_4)\cdot(\up_2\times\up_3)\right]\right\} \, .\label{UT2}
\eea
\end{subequations}
The imaginary terms in Eqs.\ (\ref{UT1}) and (\ref{UT2}) do not appear in the collision integral since
they vanish after angular integration, see Appendix \ref{appcomplex}.

\section{Angular integration}
\label{appangular}

In the first part of this appendix we compute the angular integrals in 
Eqs.\ (\ref{gammad}) and (\ref{tildegammad}). In the second part we show that the angular integral over the
imaginary terms in Eqs.\ (\ref{UT1}) and (\ref{UT2}) vanishes.

\subsection{Angular integration in Eqs.\ (\ref{gammad}) and (\ref{tildegammad})}
 
We have to compute 
\be
K\equiv \int d\Omega_1\int d\Omega_2\int d\Omega_3\int d\Omega_4 (1-\up_1\cdot\up_2)(1-\up_3\cdot\up_4)
\delta({\bf p}_1+{\bf p}_2-{\bf p}_3-{\bf p}_4) \, .
\ee
One may start with the $d\Omega_2$ integral and rewrite the $\delta$-function in terms of 
a modulus and an angular part
\be
\delta[{\bf p}_2-({\bf p}_3+{\bf p}_4-{\bf p}_1)] = \frac{1}{p_2^2}\delta(p_2-|{\bf p}_3+{\bf p}_4-{\bf p}_1|)
\delta(\Omega_2-\Omega) \, ,
\ee
where $\Omega$ stands for the angles of the vector ${\bf p}_3+{\bf p}_4-{\bf p}_1$. Thus, 
the effect of the angular $\delta$-function is to replace the direction $\up_2$ with 
$({\bf p}_3+{\bf p}_4-{\bf p}_1)/|{\bf p}_3+{\bf p}_4-{\bf p}_1|$ (the numerator being actually $p_2$
because of the radial $\delta$-function). Consequently, with ${\bf P}\equiv {\bf p}_3 + {\bf p}_4$
we obtain
\be
K = \frac{1}{p_2^2}\int d\Omega_1\int d\Omega_3\int d\Omega_4\left[1-\frac{1}{p_2}(\up_1\cdot {\bf P}-p_1)
\right](1-\up_3\cdot\up_4)\delta(p_2-|{\bf P}-{\bf p}_1|) \, .
\ee
Next, we may straightforwardly perform the $d\Omega_1$ integration by choosing the $z$-axis to 
be parallel to ${\bf P}$. As in Ref.~\cite{wadhwa} we denote
\be
p_{ij}\equiv|p_i-p_j|  \, , \qquad P_{ij} \equiv p_i+p_j \, .
\ee
Then,
\be
\int d\Omega_1\left[1-\frac{1}{p_2}(\up_1\cdot {\bf P}-p_1)
\right]\delta(p_2-|{\bf P}-{\bf p}_1|) = \frac{\pi}{p_1^2}\frac{1}{P}\Theta(P-p_{12})\Theta(P_{12}-P)
(P_{12}^2-P^2) \, .
\ee
To obtain this result, note that $\Theta(a-|b|)=\Theta(a-b)\Theta(a+b)$.
The remaining angular dependence is reduced to the angle between ${\bf p}_3$ and ${\bf p}_4$, present in $P$.
Hence, one integration, say $d\Omega_4$, is trivial while we are left with the azimuthal integral from 
$d\Omega_3$,
\bea
K&=&\frac{8\pi^3}{p_1^2p_2^2}\int_{-1}^1 dx\,\frac{1}{P}\Theta(P-p_{12})\Theta(P_{12}-P)(P_{12}^2-P^2)(1-x) 
\non
&=&\frac{4\pi^3}{p_1^2p_2^2p_3^2p_4^2}\int_{p_{34}}^{P_{34}}dP\,\Theta(P-p_{12})\Theta(P_{12}-P)(P_{12}^2-P^2)
(P_{34}^2-P^2) \, .
\eea
Since $P_{ij}>p_{ij}$, there are four orders of $p_{12},p_{34},P_{12},P_{34}$ that yield nonvanishing
integrals,
\begin{subequations}\label{orders}
\bea
p_{12}&<&p_{34}<P_{12}<P_{34} \, ,\\
p_{34}&<&p_{12}<P_{12}<P_{34} \, ,\\  
p_{34}&<&p_{12}<P_{34}<P_{12} \, , \\
p_{12}&<&p_{34}<P_{34}<P_{12} \, .
\eea
\end{subequations}
The two other possibilities, $p_{12}<P_{12}<p_{34}<P_{34}$, and $p_{34}<P_{34}<p_{12}<P_{12}$ lead 
to vanishing integrals. Consequently, 
%(keeping the terms of the sum in the same order as 
%listed in (\ref{orders})),
\be
K = \frac{4\pi^3}{p_1^2p_2^2p_3^2p_4^2}\,L(p_{12},P_{12},p_{34},P_{34}) \, ,
\ee
where we defined
\bea
L(a,b,c,d)&\equiv& \Theta(c-a)\Theta(d-b)\Theta(b-c)
\,J(c,b,b,d) +\Theta(a-c)\Theta(d-b)
\,J(a,b,b,d) \non
&&+\,\Theta(a-c)\Theta(b-d)\Theta(d-a)
\,J(a,d,b,d) + \Theta(c-a)\Theta(b-d)
\,J(c,d,b,d) \, ,
\eea     
with 
\bea
J(a,b,c,d)&\equiv&\int_a^b dP\,(c^2-P^2)(d^2-P^2) \non
&=&c^2d^2(b-a) - \frac{1}{3}(c^2+d^2)(b^3-a^3)+\frac{1}{5} (b^5-a^5) \, . 
\eea
With a simple relabeling of momenta we obtain the result for the other two terms in Eq.\ (\ref{tildegammad}),
\be
\int d\Omega_1\int d\Omega_2\int d\Omega_3\int d\Omega_4 (1+\up_1\cdot\up_3)(1+\up_2\cdot\up_4)
\delta({\bf p}_1+{\bf p}_2-{\bf p}_3-{\bf p}_4) = \frac{4\pi^3}{p_1^2p_2^2p_3^2p_4^2}\,
L(p_{24},P_{24},p_{13},P_{13})\, ,
\ee
and
\be
\int d\Omega_1\int d\Omega_2\int d\Omega_3\int d\Omega_4 (1+\up_1\cdot\up_4)(1+\up_2\cdot\up_3)
\delta({\bf p}_1+{\bf p}_2-{\bf p}_3-{\bf p}_4) = \frac{4\pi^3}{p_1^2p_2^2p_3^2p_4^2}\,
L(p_{14},P_{14},p_{23},P_{23})\, .
\ee
Consequently, with the definitions
\begin{subequations}\label{IItilde}
\bea
I(p_1,p_2,p_3,p_4)&\equiv& L(p_{12},P_{12},p_{34},P_{34}) \, , \label{I}\\
\tilde{I}(p_1,p_2,p_3,p_4)&\equiv& L(p_{12},P_{12},p_{34},P_{34}) + L(p_{24},P_{24},p_{13},P_{13})
+ L(p_{14},P_{14},p_{23},P_{23}) \, , 
\eea
\end{subequations} 
we arrive at Eqs.\ (\ref{afterang1}) and (\ref{afterang2}).

\subsection{Angular integration over imaginary terms in Eqs.\ (\ref{UT1}) and (\ref{UT2})} 
\label{appcomplex}

Here, we show that the integral  
\be
I_0\equiv \int d\Omega_1\int d\Omega_2\int d\Omega_3\int d\Omega_4 [(\up_2+\up_3)\cdot(\up_1\times\up_4)
+(\up_1+\up_4)\cdot(\up_2\times\up_3)]\delta({\bf p}_1+{\bf p}_2-{\bf p}_3-{\bf p}_4)  
\ee
is zero. After integrating over one of the angles, say over $\Omega_2$, the scalar products containing 
a cross product can all be expressed in terms of one single scalar product,
\be
I_0 = -\frac{1}{p_2^2}\left(1+\frac{p_1+p_3+p_4}{p_2}\right)\int d\Omega_1\int d\Omega_3\int d\Omega_4
\up_1\cdot(\up_3\times\up_4)\delta(p_2-|{\bf p}_3+{\bf p}_4-{\bf p}_1|) \, .
\ee
Let's consider the next integration, say over $\Omega_1$. The $\d$-function ensures that 
the distance of ${\bf p}_1$ to a fixed vector, namely ${\bf p}_3+{\bf p}_4$, equals a fixed 
value, namely $p_2$. So all allowed vectors ${\bf p}_1$ sit on a sphere around ${\bf p}_3+{\bf p}_4$
with radius $p_2$. The value of the integrand is given by the projection of ${\bf p}_1$ onto
another fixed vector, namely $\up_3\times\up_4$. Since this vector is orthogonal to ${\bf p}_3+{\bf p}_4$,
the integral over all these projections vanishes.

For the formal proof, one uses a frame with $z$-axis parallel to $\up_3\times\up_4$ and an $x$-axis
parallel to ${\bf p}_3+{\bf p}_4$ (as above, ${\bf P}\equiv{\bf p}_3+{\bf p}_4$),
\bea
\int d\Omega_1 \up_1\cdot(\up_3\times\up_4)\delta(p_2-|{\bf p}_3+{\bf p}_4-{\bf p}_1|)
&=&\int_0^\pi d\theta\,\sin\theta\cos\theta \int_0^{2\pi} d\varphi\,\delta(p_2-\sqrt{P^2+p_1^2-2Pp_1\sin\theta
\cos\varphi}) \non 
&=& \frac{2p_2}{Pp_1}\int_0^\pi d\theta\frac{\sin\theta}{|\sin\theta|}\cos\theta\left[
1-\left(\frac{p_2^2+P^2-p_1^2}{2Pp_1\sin\theta}\right)^2\right]^{-1/2} =0 \, ,
\eea
because the integrand is odd in $\theta$ over the integration range.    

\section{Rate for small temperatures} 
\label{appsmallT}

In this appendix, we determine the leading contribution for the rates 
$\Gamma_d(0,\Delta,0)$, $\Gamma_d(\Delta,0,\Delta)$, $\Gamma_d(\Delta,\Delta,\Delta)$ at small temperatures.
To this end, we analyse the following expression which originates from the Fermi distributions
in the integrands of Eqs.\ (\ref{afterang1}) and (\ref{afterang2}) (prefactors are ignored since we only 
determine the leading contribution),
\bea
\G_d(\Delta_1,\Delta_2,\Delta_4) &\sim& \sum_{e_1e_2e_4}\left(e^{e_1\sqrt{x_1^2+\varphi_1^2}}+1\right)^{-1}
\left(e^{e_2\sqrt{x_2^2+\varphi_2^2}}+1\right)^{-1} \non 
&&\times\,\left(e^{-e_1\sqrt{x_1^2+\varphi_1^2}-e_2\sqrt{x_2^2+\varphi_2^2}+e_4\sqrt{x_4^2+\varphi_4^2}-y}+1\right)^{-1}
\left(e^{-e_4\sqrt{x_4^2+\varphi_4^2}}+1\right)^{-1} \, ,
\eea
where $x_i\equiv (p_i-\mu_i)/T$, $\varphi_i\equiv \Delta(T)/T$, $y=\d\m/T$. 
This expression is integrated over $x_1$, $x_2$, $x_4$, each ranging from $-\infty$ to $\infty$ (the
lower boundary $-\infty$ actually being an approximation for $-\mu/T$).
We perform the sum and, for small $T$ (hence large $\varphi_i$), approximate
\bea
\G_d(\Delta_1,\Delta_2,\Delta_4)&\sim& 
\frac{1}{e^{\sqrt{x_4^2+\varphi_4^2}-y}+e^{\sqrt{x_1^2+\varphi_1^2}+\sqrt{x_2^2+\varphi_2^2}}}
+\frac{1}{e^{\sqrt{x_1^2+\varphi_1^2}+\sqrt{x_4^2+\varphi_4^2}-y}+e^{\sqrt{x_2^2+\varphi_2^2}}} \non
&&+\,\frac{1}{e^{\sqrt{x_2^2+\varphi_2^2}+\sqrt{x_4^2+\varphi_4^2}-y}+e^{\sqrt{x_1^2+\varphi_1^2}}} 
+\frac{1}{e^{-y}+e^{\sqrt{x_1^2+\varphi_1^2}+\sqrt{x_2^2+\varphi_2^2}+\sqrt{x_4^2+\varphi_4^2}}} \non
&&+\,\frac{1}{e^{\sqrt{x_1^2+\varphi_1^2}+\sqrt{x_2^2+\varphi_2^2}+\sqrt{x_4^2+\varphi_4^2}-y} +1}
+\frac{1}{e^{\sqrt{x_1^2+\varphi_1^2}-y}+e^{\sqrt{x_2^2+\varphi_2^2}+\sqrt{x_4^2+\varphi_4^2}}} \non
&&+\,\frac{1}{e^{\sqrt{x_2^2+\varphi_2^2}-y}+e^{\sqrt{x_1^2+\varphi_1^2}+\sqrt{x_4^2+\varphi_4^2}}}
+\frac{1}{e^{\sqrt{x_1^2+\varphi_1^2}+\sqrt{x_2^2+\varphi_2^2}-y}+e^{\sqrt{x_4^2+\varphi_4^2}}} \, .
\eea
At this point, we distinguish between the different numbers of gapped modes. First, let us discuss the 
case $\varphi_1=\varphi_2=\varphi_3\equiv\varphi$. We expand 
$\sqrt{x_i^2+\varphi^2}\simeq\varphi+x_i^2/(2\varphi)$ and introduce the new (integration) variables 
$z_i=x_i/\sqrt{2\varphi}$,
\be \label{smallT}
\G_d(\Delta,\Delta,\Delta)\sim \frac{1}{e^{\varphi-y}e^{z_1^2} + e^{2\varphi}e^{z_2^2+z_4^2}} + 
\frac{1}{e^{2\varphi-y}e^{z_1^2+z_2^2} + e^{\varphi}e^{z_4^2}} + 
\frac{1}{e^{-y} + e^{3\varphi}e^{z_1^2+z_2^2+z_4^2}} + 
\frac{1}{e^{3\varphi-y}e^{z_1^2+z_2^2+z_4^2} + 1} \, .
\ee 
Now note that for sufficiently small $T$, both $\varphi$ and $y$ have the same temperature dependence,
\be
y=\frac{\d\m}{T} \, , \qquad \varphi=\frac{\Delta(T)}{T}=
\frac{\Delta_0}{T}\sqrt{1-\left(\frac{T}{T_c}\right)^2}\to \frac{\Delta_0}{T} \, .
\ee
Therefore, the limiting behavior of $\G_d(\Delta,\Delta,\Delta)$ is different for the four 
cases $\d\m<\Delta_0$, $\Delta_0<\d\m<2\Delta_0$, $2\Delta_0<\d\m<3\Delta_0$, and $3\Delta_0<\d\m$. 
In particular, the latter case leads to a nonzero value of $\G_d$ for $T\to 0$. 
This can be seen from Eq.\ (\ref{smallT}), where the 1 in the denominator of the fourth term on the 
right-hand side survives for $3\Delta_0<\d\m$. We show the results for all four cases in the first row 
of Table \ref{tablesmallT}.

\begin{table}  
\begin{tabular}[t]{|c||c|c|c|c|}
\hline
 & $\;\;$ $0<\d\m<\Delta_0$$\;\;$  & $\;\;$$\Delta_0<\d\m<2\Delta_0$$\;\;$ & $\;\;$
$2\Delta_0<\d\m<3\Delta_0$$\;\;$ & $\;\;$$3\Delta_0<\d\m$ $\;\;$
\\ \hline\hline
$\;\;$$\G_d(\Delta,\Delta,\Delta)$$\;\;$  & $e^{-(2\Delta-\d\m)/T}$ & $e^{-\Delta/T}$ & $e^{-(3\Delta-\d\m)/T}$ & 1
\\ \hline
$\G_d(\Delta,0,\Delta)$  & $e^{-\Delta/T}$ & $e^{-(2\Delta-\d\m)/T}$ & 1 & 1
\\ \hline
$\G_d(0,\Delta,0)$  & $e^{-(\Delta-\d\m)/T}$ & 1 & 1 & 1
\\ \hline
$\G_d(0,0,0)$  & 1 & 1 & 1 & 1 
\\ \hline
\end{tabular}
\caption{Leading small temperature behavior of the rates with three, two, one, and zero gapped modes. A ``1''
indicates that the rate does not vanish at $T=0$.}
\label{tablesmallT}
\end{table}

Next, we consider the case $\varphi_1=\varphi_4\equiv\varphi$, $\varphi_2=0$. 
The analogue to Eq.\ (\ref{smallT}) is
\bea
\G_d(\Delta,0,\Delta)&\sim& \frac{1}{e^{\varphi-y}e^{z_4^2} + e^{\varphi}e^{z_1^2+x_2}} + 
\frac{1}{e^{2\varphi-y}e^{z_1^2+z_4^2} + e^{x_2}} + 
\frac{1}{e^{\varphi-y}e^{x_2+z_4^2} + e^{\varphi}e^{z_1^2}} \non
&&+\, \frac{1}{e^{-y} + e^{2\varphi}e^{z_1^2+x_2+z_4^2}} 
+\frac{1}{e^{2\varphi-y}e^{z_1^2+x_2+z_4^2} + 1} 
+\frac{1}{e^{-y}e^{x_2} + e^{2\varphi}e^{z_1^2+z_4^2}} \, .
\eea 
Again, the results for different values of $\d\m$ are shown in Table \ref{tablesmallT}.

Finally, for $\varphi_1=\varphi_4=0$, $\varphi_2\equiv\varphi$ we have
\bea
\G_d(0,\Delta,0)&\sim& \frac{1}{e^{-y}e^{x_4} + e^{\varphi}e^{x_1+z_2^2}} + 
\frac{1}{e^{-y}e^{x_1+x_4} + e^{\varphi}e^{z_2^2}} + 
\frac{1}{e^{\varphi-y}e^{z_2^2+x_4} + e^{x_1}} \non
&&+\,\frac{1}{e^{-y} + e^{\varphi}e^{x_1+z_2^2+x_4}} 
+ \frac{1}{e^{\varphi-y}e^{x_1+z_2^2+x_4} + 1} 
+\frac{1}{e^{\varphi-y}e^{z_2^2} + e^{x_1+x_4}} \, .
\eea 
Here, only for $\d\m < \Delta_0$, the rate goes to zero for $T\to 0$. The result is shown in the third
row of Table \ref{tablesmallT}. For completeness, we added the trivial result that for $\G_d(0,0,0)$
the rate is nonzero at $T=0$ for all $\d\m>0$, see fourth row in Table \ref{tablesmallT}.


\begin{thebibliography}{99}

\bibitem{perry}
J.~C.~Collins and M.~J.~Perry,
%``Superdense Matter: Neutrons Or Asymptotically Free Quarks?,''
Phys.\ Rev.\ Lett.\  {\bf 34}, 1353 (1975).
%%CITATION = PRLTA,34,1353;%%

%\cite{Bailin:1983bm}
\bibitem{bailin}
  D.~Bailin and A.~Love,
  %``Superfluidity And Superconductivity In Relativistic Fermion Systems,''
  Phys.\ Rept.\  {\bf 107}, 325 (1984).

\bibitem{bcs} 
J.\ Bardeen, L.N.\ Cooper, and J.R.\ Schrieffer, 
Phys.\ Rev.\ {\bf 108}, 1175 (1957).


\bibitem{reviews}
For reviews, see
  %%CITATION = PRPLC,107,325;%%
K.~Rajagopal and F.~Wilczek,
%``The condensed matter physics of QCD,''
arXiv:hep-ph/0011333;
%%CITATION = HEP-PH 0011333;%%
%%\bibitem{Alford:2001dt}
M.~G.~Alford,
%``Color superconducting quark matter,''
Ann.\ Rev.\ Nucl.\ Part.\ Sci.\  {\bf 51}, 131 (2001)
[arXiv:hep-ph/0102047];
%cite{Nardulli:2002ma}
%%\bibitem{Nardulli:2002ma}
G.~Nardulli,
%``Effective description of QCD at very high densities,''
Riv.\ Nuovo Cim.\  {\bf 25N3}, 1 (2002)
[arXiv:hep-ph/0202037];
%%CITATION = HEP-PH 0202037;%%
%\cite{Reddy:2002ri}
%%\bibitem{Reddy:2002ri}
S.~Reddy,
%``Novel phases at high density and their roles in the structure
%and evolution of neutron stars,''
Acta Phys.\ Polon.\ B {\bf 33}, 4101 (2002)
[arXiv:nucl-th/0211045];
%%CITATION = NUCL-TH 0211045;%%
%\cite{Schafer:2003vz}
%%\bibitem{Schafer:2003vz}
T.~Sch\"afer,
%``Quark matter,''
arXiv:hep-ph/0304281;
D.~H.~Rischke, Prog.\ Part.\ Nucl.\ Phys.\ {\bf 52}, 197 (2004) 
[arXiv:nucl-th/0305030];
%\cite{Alford:2003eg}
%\bibitem{Alford:2003eg}
  M.~Alford,
  %``Dense quark matter in nature,''
  Prog.\ Theor.\ Phys.\ Suppl.\  {\bf 153}, 1 (2004)
  [arXiv:nucl-th/0312007];
  %%CITATION = NUCL-TH 0312007;%%
%\cite{Alford:2003eg}
%%CITATION = NUCL-TH 0312007;%%
%\cite{Buballa:2003qv}
%\bibitem{Buballa:2003qv}
  M.~Buballa,
  %``NJL model analysis of quark matter at large density,''
  Phys.\ Rept.\  {\bf 407}, 205 (2005)
  [arXiv:hep-ph/0402234];
  %%CITATION = HEP-PH 0402234;%%
%\cite{Ren:2004nn}
%\bibitem{Ren:2004nn}
  H.~c.~Ren,
  %``Color superconductivity of QCD at high baryon density,''
  arXiv:hep-ph/0404074;
  %%CITATION = HEP-PH 0404074;%%
%\bibitem{Huang:2004ik}
  M.~Huang,
  %``Color superconductivity at moderate baryon density,''
  Int.\ J.\ Mod.\ Phys.\ E {\bf 14}, 675 (2005)
  [arXiv:hep-ph/0409167];
   I.~A.~Shovkovy,
  %``Two lectures on color superconductivity,''
  Found.\ Phys.\  {\bf 35}, 1309 (2005)
  [arXiv:nucl-th/0410091]. 
  T.~Sch\"afer,
  %``Phases of QCD,''
  arXiv:hep-ph/0509068.


\bibitem{Alford:1998mk}
M.~G.~Alford, K.~Rajagopal and F.~Wilczek,
%``Color-flavor locking and chiral symmetry breaking in high density {QCD},''
Nucl.\ Phys.\ B {\bf 537}, 443 (1999)
[arXiv:hep-ph/9804403].
%%CITATION = HEP-PH 9804403;%%

\bibitem{Rajagopal:2005dg}
  K.~Rajagopal and A.~Schmitt,
  %``Stressed pairing in conventional color superconductors is unavoidable,''
  Phys.\ Rev.\ D {\bf 73}, 045003 (2006)
  [arXiv:hep-ph/0512043].

\bibitem{spin1}
T.~Sch\"afer,
%``Quark hadron continuity in QCD with one flavor,''
Phys.\ Rev.\ D {\bf 62}, 094007 (2000)
[arXiv:hep-ph/0006034];
%%CITATION = HEP-PH 0006034;%%
M.~Buballa, J.~Hosek and M.~Oertel,
%``Anisotropic admixture in color-superconducting quark matter,''
Phys.\ Rev.\ Lett.\  {\bf 90}, 182002 (2003)
[arXiv:hep-ph/0204275];
%%CITATION = HEP-PH 0204275;%%
A.~Schmitt, Q.~Wang and D.~H.~Rischke,
%``When the transition temperature in color superconductors is not like in BCS
%theory,''
Phys.\ Rev.\ D {\bf 66}, 114010 (2002)
[arXiv:nucl-th/0209050];
%%CITATION = NUCL-TH 0209050;%%
M.~G.~Alford, J.~A.~Bowers, J.~M.~Cheyne and G.~A.~Cowan,
%``Single color and single flavor color superconductivity,''
Phys.\ Rev.\ D {\bf 67}, 054018 (2003)
[arXiv:hep-ph/0210106];
%%CITATION = HEP-PH 0210106;%%
A.~Schmitt, {\it Ph.D.\ dissertation},
%``Spin-one Color Superconductivity in Cold and Dense Quark Matter,''
arXiv:nucl-th/0405076;
%%CITATION = NUCL-TH 0405076;%%
%\bibitem{Schmitt:2004et}
  A.~Schmitt,
  %``The ground state in a spin-one color superconductor,''
  Phys.\ Rev.\ D {\bf 71}, 054016 (2005)
  [arXiv:nucl-th/0412033].
  %%CITATION = NUCL-TH 0412033;%%

\bibitem{LOFF}
M.~G.~Alford, J.~A.~Bowers and K.~Rajagopal,
%``Crystalline color superconductivity,''
Phys.\ Rev.\ D {\bf 63}, 074016 (2001)
[arXiv:hep-ph/0008208];
  I.~Giannakis and H.~C.~Ren,
  %``Chromomagnetic instability and the LOFF state in a two flavor color
  %superconductor,''
  Phys.\ Lett.\ B {\bf 611}, 137 (2005)
  [arXiv:hep-ph/0412015];
  R.~Casalbuoni, R.~Gatto, N.~Ippolito, G.~Nardulli and M.~Ruggieri,
  %``Ginzburg-Landau approach to the three flavor LOFF phase of QCD,''
  Phys.\ Lett.\ B {\bf 627}, 89 (2005)
  [Erratum-ibid.\ B {\bf 634}, 565 (2006)]
  [arXiv:hep-ph/0507247];
  M.~Mannarelli, K.~Rajagopal and R.~Sharma,
  % ``Testing the Ginzburg-Landau approximation for three-flavor crystalline
  % color superconductivity,''
  Phys.\ Rev.\ D {\bf 73}, 114012 (2006)
  [arXiv:hep-ph/0603076];
  %\bibitem{Rajagopal:2006dp}
  K.~Rajagopal and R.~Sharma,
  %``The crystallography of strange quark matter,''
  arXiv:hep-ph/0606066.

\bibitem{Alford:1999pb}
  M.~G.~Alford, J.~Berges and K.~Rajagopal,
  %``Magnetic fields within color superconducting neutron star cores,''
  Nucl.\ Phys.\ B {\bf 571}, 269 (2000)
  [arXiv:hep-ph/9910254];
%\bibitem{Schmitt:2003xq}
  A.~Schmitt, Q.~Wang and D.~H.~Rischke,
  %``Electromagnetic Meissner effect in spin-one color superconductors,''
  Phys.\ Rev.\ Lett.\  {\bf 91}, 242301 (2003)
  [arXiv:nucl-th/0301090];
%\bibitem{Schmitt:2003aa}
  A.~Schmitt, Q.~Wang and D.~H.~Rischke,
  %``Mixing and screening of photons and gluons in a color superconductor,''
  Phys.\ Rev.\ D {\bf 69}, 094017 (2004)
  [arXiv:nucl-th/0311006].

\bibitem{Ferrer:2005vd}
  E.~J.~Ferrer, V.~de la Incera and C.~Manuel,
  %``Magnetic color flavor locking phase in high density QCD,''
  Phys.\ Rev.\ Lett.\  {\bf 95}, 152002 (2005)
  [arXiv:hep-ph/0503162];
%\bibitem{Ferrer:2006vw}
  E.~J.~Ferrer, V.~de la Incera and C.~Manuel,
  % ``Color-superconducting gap in the presence of a magnetic field,''
  Nucl.\ Phys.\ B {\bf 747}, 88 (2006)
  [arXiv:hep-ph/0603233].

\bibitem{Shovkovy:2002kv}
  I.~A.~Shovkovy and P.~J.~Ellis,
  %``Thermal conductivity of dense quark matter and cooling of stars,''
  Phys.\ Rev.\ C {\bf 66}, 015802 (2002)
  [arXiv:hep-ph/0204132].

\bibitem{Carter:2000xf}
  G.~W.~Carter and S.~Reddy,
  %``Neutrino propagation in color superconducting quark matter,''
  Phys.\ Rev.\ D {\bf 62}, 103002 (2000)
  [arXiv:hep-ph/0005228];
%\bibitem{Jaikumar:2002vg}
  P.~Jaikumar, M.~Prakash and T.~Sch\"afer,
  %``Neutrino emission from Goldstone modes in dense quark matter,''
  Phys.\ Rev.\ D {\bf 66}, 063003 (2002)
  [arXiv:astro-ph/0203088];
%\bibitem{Reddy:2002xc}
  S.~Reddy, M.~Sadzikowski and M.~Tachibana,
  %``Neutrino rates in color flavor locked quark matter,''
  Nucl.\ Phys.\ A {\bf 714}, 337 (2003)
  [arXiv:nucl-th/0203011];
%\bibitem{Kundu:2004mz}
  J.~Kundu and S.~Reddy,
  % ``Neutrino Scattering Off Pair-Breaking And Collective Excitations In
  %Superfluid Neutron Matter And In Color-Flavor Locked Quark Matter,''
  Phys.\ Rev.\ C {\bf 70}, 055803 (2004)
  [arXiv:nucl-th/0405055];
%\bibitem{Anglani:2006br}
  R.~Anglani, M.~Mannarelli, G.~Nardulli and M.~Ruggieri,
  % ``Neutrino emission from compact stars and inhomogeneous color
  %superconductivity,''
  arXiv:hep-ph/0607341.

\bibitem{Jaikumar:2005hy}
  P.~Jaikumar, C.~D.~Roberts and A.~Sedrakian,
  %``Direct Urca neutrino rate in colour superconducting quark matter,''
  Phys.\ Rev.\ C {\bf 73}, 042801 (2006)
  [arXiv:nucl-th/0509093].

\bibitem{Schmitt:2005wg}
  A.~Schmitt, I.~A.~Shovkovy and Q.~Wang,
  %``Neutrino emission and cooling rates of spin-one color superconductors,''
  Phys.\ Rev.\ D {\bf 73}, 034012 (2006)
  [arXiv:hep-ph/0510347].

\bibitem{Wang:2006tg}
  Q.~Wang, Z.~g.~Wang and J.~Wu,
  %``Phase space and quark mass effects in neutrino emissions in a color
  %superconductor,''
  arXiv:hep-ph/0605092. 

\bibitem{Sa'd:2006qv}
  B.~A.~Sa'd, I.~A.~Shovkovy and D.~H.~Rischke,
%   ``Bulk viscosity of spin-one color superconductors I: Two-flavor quark
  %matter,''
  arXiv:astro-ph/0607643.

\bibitem{Drago:2003wg}
  A.~Drago, A.~Lavagno and G.~Pagliara,
  %``Bulk viscosity in hybrid stars,''
  Phys.\ Rev.\ D {\bf 71}, 103004 (2005)
  [arXiv:astro-ph/0312009].

\bibitem{Alford:2002kj}
M.~Alford and K.~Rajagopal,
%``Absence of two-flavor color superconductivity in compact stars,''
JHEP {\bf 0206}, 031 (2002)
[arXiv:hep-ph/0204001].
%%CITATION = HEP-PH 0204001;%%

\bibitem{Steiner:2002gx}
A.~W.~Steiner, S.~Reddy and M.~Prakash,
%``Color-neutral superconducting quark matter,''
Phys.\ Rev.\ D {\bf 66}, 094007 (2002)
[arXiv:hep-ph/0205201].
%%CITATION = HEP-PH 0205201;%%


\bibitem{Ruster:2005jc}
  S.~B.~R\"uster, V.~Werth, M.~Buballa, I.~A.~Shovkovy and D.~H.~Rischke,
  %``The phase diagram of neutral quark matter: Self-consistent treatment of
  %quark masses,''
  Phys.\ Rev.\ D {\bf 72}, 034004 (2005)
  [arXiv:hep-ph/0503184].
  %%CITATION = HEP-PH 0503184;%%

\bibitem{Abuki:2005ms}
  H.~Abuki and T.~Kunihiro,
 %  ``Extensive study of phase diagram for charge neutral homogeneous quark
 %  matter affected by dynamical chiral condensation: Unified picture for
 % thermal unpairing transitions from weak to strong coupling,''
  Nucl.\ Phys.\ A {\bf 768}, 118 (2006)
  [arXiv:hep-ph/0509172].
  %%CITATION = HEP-PH 0509172;%%

\bibitem{Manuel:2004iv}
  C.~Manuel, A.~Dobado and F.~J.~Llanes-Estrada,
  %``Shear viscosity in a CFL quark star,''
  JHEP {\bf 0509}, 076 (2005)
  [arXiv:hep-ph/0406058].

\bibitem{Kokkotas:2001ze}
  K.~D.~Kokkotas and N.~Andersson,
  %``Oscillation and instabilities of relativistic stars,''
  arXiv:gr-qc/0109054.
  %%CITATION = GR-QC 0109054;%%

\bibitem{Andersson:1997xt}
  N.~Andersson,
  %``A new class of unstable modes of rotating relativistic stars,''
  Astrophys.\ J.\  {\bf 502}, 708 (1998)
  [arXiv:gr-qc/9706075];

\bibitem{Andersson:2001bz}
  N.~Andersson and G.~L.~Comer,
  %``On the dynamics of superfluid neutron star cores,''
  Mon.\ Not.\ Roy.\ Astron.\ Soc.\  {\bf 328}, 1129 (2001)
  [arXiv:astro-ph/0101193].
  %%CITATION = ASTRO-PH 0101193;%%

\bibitem{Lindblom:2000jw}
For a lecture on r-mode instabilities see
  L.~Lindblom,
  %``Neutron star pulsations and instabilities,''
  arXiv:astro-ph/0101136.

\bibitem{Weber:2004kj}
  F.~Weber,
  %``Strange quark matter and compact stars,''
  Prog.\ Part.\ Nucl.\ Phys.\  {\bf 54}, 193 (2005)
  [arXiv:astro-ph/0407155];
  %%CITATION = ASTRO-PH 0407155;%
%\bibitem{Weber:2006iw}
  F.~Weber, A.~Torres i Cuadrat, A.~Ho and P.~Rosenfield,
  %``Strangeness in compact stars,''
  PoS {\bf JHW2005}, 018 (2006)
  [arXiv:astro-ph/0602047].
  %%CITATION = ASTRO-PH 0602047;%%

\bibitem{cutler}
C.\ Cutler and L.\ Lindblom,
%The effect of viscosity on neutron star oscillations
Astrophys.\ J.\ {\bf 314}, 234 (1987);
%\bibitem{Sawyer:1989dp}
  R.~F.~Sawyer,
  %``Bulk Viscosity Of Hot Neutron--Star Matter And The Maximum Rotation Rates
  %Of Neutron Stars,''
  Phys.\ Rev.\ D {\bf 39}, 3804 (1989).

\bibitem{Wang:1985tg}
  Q.~D.~Wang and T.~Lu,
  %``The Damping Effects Of The Vibrations In The Core Of A Neutron Star,''
  Phys.\ Lett.\ B {\bf 148}, 211 (1984).
  %%CITATION = PHLTA,B148,211;%%

\bibitem{Arnold:2006fz}
  P.~Arnold, C.~Dogan and G.~D.~Moore,
  %``The bulk viscosity of high-temperature QCD,''
  arXiv:hep-ph/0608012.
  %%CITATION = HEP-PH 0608012;%%

\bibitem{bhatta}
P.\ Bhattacharyya, C.J.\ Pethick and H.\ Smith, Phys.\ Rev.\ B {\bf 15}, 3367 (1977);
%Transport and relaxation processes in superfluid $^3$He close to the transition temperature
C.J.\ Pethick, P.\ Bhattacharyya  and H.\ Smith, Phys.\ Rev.\ B {\bf 15}, 3384 (1977).
%Transport properties in superfluid $^3$He-B at low temperatures

\bibitem{vollhardt}
D.\ Vollhardt and P.\ W\"olfle, 
{\it The Superfluid Phases of Helium 3} (Taylor \& Francis, London, 1990).  

\bibitem{madsen}
  J.~Madsen,
  %``Bulk viscosity of strange quark matter, damping of quark star vibration, and
  %the maximum rotation rate of pulsars,''
  Phys.\ Rev.\ D {\bf 46}, 3290 (1992).

\bibitem{anand}
  J.~D.~Anand, N.~Chandrika Devi, V.~K.~Gupta and S.~Singh,
  %``Bulk viscosity of strange quark matter in density dependent quark mass
  %model,''
  Pramana {\bf 54}, 737 (2000).

\bibitem{Haensel:2000vz}
  P.~Haensel, K.~P.~Levenfish and D.~G.~Yakovlev,
  %``Bulk viscosity in superfluid neutron star cores. I. Direct Urca processes
  %in npe\mu matter,''
  Astron.\ Astrophys.\ {\bf 357}, 1157 (2000) [arXiv:astro-ph/0004183].

\bibitem{Lindblom:2001hd}
  L.~Lindblom and B.~J.~Owen,
  %``Effect of hyperon bulk viscosity on neutron-star r-modes,''
  Phys.\ Rev.\ D {\bf 65}, 063006 (2002)
  [arXiv:astro-ph/0110558].

\bibitem{KB} L.~P.~Kadanoff and G.~Baym,
{\it Quantum Statistical Mechanics}
(Benjamin, New York, 1962).

\bibitem{friman}
M.~Sch\"onhofen, M.~Cubero, B.~L.~Friman, W.~N\"orenberg and G.~Wolf,
%``Covariant kinetic equations and relaxation processes in relativistic heavy
%ion collisions,''
Nucl.\ Phys.\ A {\bf 572}, 112 (1994).
%%CITATION = NUPHA,A572,112;%%

\bibitem{sedrakian}
A.~Sedrakian and A.~Dieperink,
%``Coherence effects and neutrino pair bremsstrahlung in neutron stars,''
Phys.\ Lett.\ B {\bf 463}, 145 (1999) [arXiv:nucl-th/9905039].
%%CITATION = NUCL-TH 9905039;%%

\bibitem{wadhwa}
A.\ Wadhwa, V.K.\ Gupta, S.\ Singh, and J.D.\ Anand, J.\ Phys. G {\bf 9}, 1137 (1995).
%On the cooling of neutron stars

\bibitem{Madsen:1993xx}
  J.~Madsen,
  %``Rate of the weak reaction s + u $\to$ u + d in quark matter,''
  Phys.\ Rev.\ D {\bf 47}, 325 (1993).

\bibitem{Madsen:1999ci}
  J.~Madsen,
  %``Probing strange stars and color superconductivity by r-mode  instabilities
  %in millisecond pulsars,''
  Phys.\ Rev.\ Lett.\  {\bf 85}, 10 (2000)
  [arXiv:astro-ph/9912418].

\bibitem{Ruster:2005ib}
  S.~B.~Ruster, V.~Werth, M.~Buballa, I.~A.~Shovkovy and D.~H.~Rischke,
  % ``The phase diagram of neutral quark matter: The effect of neutrino
  %trapping,''
  Phys.\ Rev.\ D {\bf 73}, 034025 (2006)
  [arXiv:hep-ph/0509073].

\bibitem{Lindblom:2000az}
  L.~Lindblom, J.~E.~Tohline and M.~Vallisneri,
  %``Non-Linear Evolution of the r-Modes in Neutron Stars,''
  Phys.\ Rev.\ Lett.\  {\bf 86}, 1152 (2001)
  [arXiv:astro-ph/0010653].
  %%CITATION = ASTRO-PH 0010653;%%

\bibitem{Sahu:2001iv}
  P.~K.~Sahu, G.~F.~Burgio and M.~Baldo,
  %``Radial modes of neutron stars with a quark core,''
  Astrophys.\ J.\  {\bf 566}, L89 (2002)
  [arXiv:astro-ph/0111414];
  %%CITATION = ASTRO-PH 0111414;
%\bibitem{Haensel:2002qw}
  P.~Haensel, K.~P.~Levenfish and D.~G.~Yakovlev,
  %``Adiabatic Index of Dense Matter and Damping of Neutron Star Pulsations,''
  Astron.\ Astrophys.\  {\bf 394}, 213 (2002)
  [arXiv:astro-ph/0208078];
  %%CITATION = ASTRO-PH 0208078;%%
%\bibitem{Gusakov:2005dz}
  M.~E.~Gusakov, D.~G.~Yakovlev and O.~Y.~Gnedin,
  %``Thermal evolution of a pulsating neutron star,''
  Mon.\ Not.\ Roy.\ Astron.\ Soc.\  {\bf 361}, 1415 (2005)
  [arXiv:astro-ph/0502583].
  %%CITATION = ASTRO-PH 0502583;%%

\bibitem{preparation}
M.\ Alford, M.\ Braby, S.\ Reddy, T.\ Sch\"afer, in preparation.


\end{thebibliography}
\end{document}